\pdfoutput=1

\documentclass[conference,10pt]{IEEEtran}

\ifCLASSINFOpdf
   \usepackage[pdftex]{graphicx}
   \graphicspath{{./}{./img/}}
   \DeclareGraphicsExtensions{.pdf,.jpeg,.jpg,.png,.eps,.ps}
\else
  \usepackage[dvips]{graphicx}
  \graphicspath{{./}{./img/}{./eps/}}
  \DeclareGraphicsExtensions{.pdf,.jpeg,.jpg,.png,.eps,.ps}
\fi
\usepackage{cite}
\usepackage[cmex10]{amsmath}
\usepackage{algorithmic}
\usepackage{array}
\usepackage{mdwmath}
\usepackage{mdwtab}
\usepackage{eqparbox}
\usepackage[font=footnotesize]{subfig}
\usepackage{stfloats}
\usepackage{url}

\usepackage[lined, ruled, linesnumbered,noend]{algorithm2e}
\usepackage[bookmarks=false]{hyperref}
\usepackage{latexsym}
\usepackage{amsfonts,amssymb,amsthm}
\usepackage[usenames,dvipsnames,svgnames,table]{xcolor}
\let\labelindent\relax
\usepackage{enumitem}
\usepackage{pbox}
\usepackage{mathtools}
\usepackage{float}
\usepackage{pgffor}
\usepackage{bm}		%
\usepackage{tabularx}	%
\usepackage{framed}	
\usepackage{cleveref}	%
\usepackage{pifont}		%
\usepackage[top=0.763in, bottom=0.68in, left=0.68in, right=0.6in]{geometry}
\IEEEoverridecommandlockouts

\IEEEaftertitletext{\vspace{2mm}}
\newtheorem{definition}{\bf Definition}	%
\newtheorem{theorem}{\bf Theorem}		%
\newcounter{appdx}
\newcommand{\kETAL}     {{\em et~al.}}		%
\newcommand{\kIE}     {{\em i.e.}}		%

\newcommand{\smallfont}  {}

\definecolor{dgreen}{rgb}{0,0.655,0.149}

\definecolor{purple}{HTML}{7F00FF}

\newcommand{\redc}[1]{{\color{red}}}

\SetKw{Continue}{continue}
\SetKw{Break}{break}

\makeatletter
\newcommand{\nosemic}{\renewcommand{\@endalgocfline}{\relax}}%
\newcommand{\dosemic}{\renewcommand{\@endalgocfline}{\algocf@endline}}%
\let\oldnl\nl%
\newcommand{\nonl}{\renewcommand{\nl}{\let\nl\oldnl}}%
\makeatother

\SetKwInOut{Init}{Initialize}
\SetKwInOut{OptIn}{Optional input}

\begin{document}
\title{ESDI: Entanglement Scheduling and Distribution in the
Quantum Internet\vspace{-1em}

\author{\IEEEauthorblockN{Huayue~Gu, Ruozhou~Yu, Zhouyu~Li, Xiaojian~Wang, Fangtong~Zhou}}
\IEEEcompsocitemizethanks{\IEEEcompsocthanksitem
\vspace{-1em}
\IEEEcompsocthanksitem Gu, Yu, Li, Wang and Zhou (\{hgu5, ryu5, zli85, xwang244, fzhou\}@ncsu.edu) are all with NC State University, Raleigh, NC 27606, USA. }
}
\maketitle

\begin{abstract}
Quantum entanglement distribution between remote nodes is key to many promising quantum applications.
Existing mechanisms have mainly focused on improving throughput and fidelity via entanglement routing or single-node scheduling.
This paper considers entanglement scheduling and distribution among many source-destination pairs with different requests over an entire quantum network topology. 
Two practical scenarios are considered. When requests do not have deadlines, we seek to minimize the average completion time of the communication requests. If deadlines are specified, we seek to maximize the number of requests whose deadlines are met.
Inspired by optimal scheduling disciplines in conventional single-queue scenarios, we design a general optimization framework for entanglement scheduling and distribution called {ESDI}, and develop a probabilistic protocol to implement the optimized solutions in a general buffered quantum network.
We develop a discrete-time quantum network simulator for evaluation.
Results show the superior performance of ESDI compared to existing solutions.
\end{abstract}

\begin{IEEEkeywords}
Quantum network, entanglement scheduling, entanglement routing
\end{IEEEkeywords}

\section{Introduction}
\label{sec:intro}
\noindent 
A quantum network enables efficient and secure quantum communication based on quantum entanglement~\cite{koashi1998no}. 
Long-distance quantum communication is key to various novel quantum applications including quantum key distribution (QKD)~\cite{bennett2020quantum}, distributed quantum computing (DQC)~\cite{cacciapuoti2019quantum,cicconetti2022resource} and quantum cryptography \cite{sriram_globecom}. 
Recently, practical quantum networks have been built around the world~\cite{yin2017satellite,dahlberg2019link}, such as the DARPA quantum network~\cite{elliott2002building}, SECOQC Vienna QKD network~\cite{peev2009secoqc}, and the Tokyo QKD network~\cite{sasaki2011field}. 

Entanglement is the most crucial resource in a quantum network. 
In quantum information processing and communication, information is represented as \emph{quantum bits} (called \emph{qubits}), which cannot be transmitted via classical communication channels without information loss.
To transmit quantum information between two arbitrary quantum-capable devices, entanglement must be leveraged for the execution of quantum protocols.
The primary function of a quantum network is to distribute entanglements between nodes over long distances.
Because of its importance, entanglement distribution has received significant attention recently. Prior work has focused on entanglement routing, \emph{i.e.}, finding paths to establish end-to-end entanglements with high throughput and/or quality~\cite{zhao2021redundant,zhao2022e2e,zeng2022multi}.
In this line of research, the most common goal is to maximize network-wide throughput of entanglement distribution for all source-destination (SD) pairs, potentially with constraints on the quality of the distributed entanglements.
In reality, however, each SD pair may utilize the quantum network to support specific applications, and different applications have different requirements for entanglement distribution rate and time constraints.
Considering two typical quantum applications, QKD and DQC. QKD relies on a \emph{long-term stream} of entanglements between two parties requesting secure communications, and is not sensitive to instantaneous entanglement distribution rate, as long as sufficient entanglements are accumulated over a relatively long period of time. Meanwhile, DQC has a stringent requirement for completing communication tasks as quickly as possible, to avoid information decoherence in quantum memories.

The above motivates scheduling entanglements among SD pairs while considering different communication requirements and demands. Existing work has formulated entanglement scheduling in simple queueing scenarios, focusing on scheduling at a single quantum switch~\cite{panigrahy2022capacity}. The assumption that all nodes should connect to a central quantum switch is rather strong and unrealistic. Given that real-world quantum links, such as optical fiber, can only distribute entanglements over no more than a few hundred kilometers~\cite{neumann2022continuous}, a general multi-hop network topology is realistic in building large-scale quantum networks~\cite{zeng2022multi}. Extending existing scheduling algorithms to this multi-hop scenario is very challenging, as scheduling in a queueing network is generally NP-hard even without any quantum-specific characteristic being considered~\cite{shah2012randomized}.

In this paper, we study scheduling in a general quantum network, while considering  SD pairs with different requests (called \emph{commodities}) for quantum communication demands and/or deadlines.
Two scenarios are specifically considered: 1) each commodity has a demand (number of entanglements) but not a deadline, in which case we seek to minimize the average completion time for all commodities' demands; 2) each commodity has a demand and a deadline, in which case we seek to maximize the number of commodities whose deadlines are met.
In addition to hardness of the scheduling problem, challenges come from the probabilistic nature of quantum operations in entanglement distribution, making it impossible to obtain an exact estimation of the completion time of each commodity.
Our main contributions are as follows:
\begin{itemize}
    \item We study and formally define the entanglement scheduling and distribution problem in a general quantum network with heterogeneous quantum applications. 
    \item We propose \textbf{ESDI}, a general framework for \emph{entanglement scheduling and distribution}, and propose two entanglement scheduling and distribution algorithms: ESDI-O for commodities having demands but no deadlines, and ESDI-E for commodities having demands and deadlines.
    \item We develop practical probabilistic protocols for  entanglement distribution in a buffered quantum network.
    \item Extensive simulation results show the superior performance of our solutions compared to state-of-the-arts. %
\end{itemize}

\noindent\textbf{Organization:}
\S\ref{sec:rw} introduces the background.
\S\ref{sec:model} presents our quantum network model.
\S\ref{sec:basic} proposes a general framework for multi-commodity entanglement scheduling and distribution.
\S\ref{sec:sjf} and \S\ref{sec:edf} propose algorithms for scheduling commodities without and with deadlines, respectively.
\S\ref{sec:esdi} presents our protocol design.
\S\ref{sec:results} and \S\ref{sec:conclusions} present simulation results and conclusion, respectively.

\section{Background and Related Work}
\label{sec:rw}
\noindent
Quantum communication transfers quantum states from one place to another.
A quantum network enables long-distance quantum communication between arbitrary end points~\cite{singh2021quantum}.

Early work mainly focused on idealized network topologies including square-grid~\cite{pant2019routing}, ring and sphere topologies~\cite{schoute2016shortcuts}.
Unfortunately, these ideal network designs are far from realistic due to physical and geographical limitations.
Considering a general network topology, recent works have paid much attention to entanglement routing for multiple SD pairs~\cite{Dai2020a}. 
Shi~\kETAL{}~\cite{shi2020concurrent} proposed two algorithms, Q-PATH and Q-CAST, for entanglement routing to maximize the throughput (entanglement distribution rate).
Zhao~\kETAL{}~\cite{zhao2021redundant} designed a redundant entanglement provisioning and selection (REPS) algorithm to maximize throughput for multiple SD pairs in a circuit-switched, multi-hop quantum network. 
Zeng~\kETAL{}~\cite{zeng2022multi} proposed an integer programming-based solution with branch-and-price to maximize throughput.
Farahbakhsh~\kETAL{}~\cite{farahbakhsh2022opportunistic} presented opportunistic routing with theoretical analysis.
None of the above studies have considered the scheduling of entanglement distribution in a general quantum network. 
Panigrahy~\kETAL{}~\cite{panigrahy2022capacity} developed a max-weight scheduling policy to stabilize queues in a star-shaped quantum network. 
Pouryousef~\kETAL{}~\cite{pouryousef2022quantum} studied the link entanglement and storage resource allocation problem in a quantum overlay network.
These works focus on scheduling in a specialized, single-repeater quantum network.
Li~\kETAL{}~\cite{li2021effective} proposed an effective routing scheme for multiple requests from SD pairs, but they only focus on the fair sharing of entanglements without considering the characteristics of requests.
Meanwhile, most research in entanglement routing considers a \emph{bufferless quantum network}, where entanglement generation and swapping must be completed in one time slot. Otherwise, the intermediate entanglements will be discarded.
With recent advances in quantum memories with long coherence times, quantum networks with buffers have been shown to improve quantum network performance and throughput~\cite{Dai2020a, gu2023fendi}.
Dai~\kETAL{}~\cite{Dai2020a} first studied the optimal remote entanglement distribution (ORED) protocol in a buffered quantum network for a single SD pair and provided an upper throughput bound for any entanglement routing protocols.
Dai~\kETAL{}~\cite{quantum-queueing-delay} also analyzed the queuing delay for a single channel in a quantum network.
However, the ORED protocol only considers one SD pair in a quantum network and may not be practical for large-scale operations. 
This paper considers multiple SD pairs with heterogeneous quantum communication requests, and designs a comprehensive framework to satisfy their demands.

\begin{table}
\caption{Notation Table}
\label{notation}
\footnotesize
\begin{tabular}{p{2.5cm}p{5.5cm}}
\hline
Parameters & Description\\
\hline
$G=(V, E)$ & quantum network with nodes $V$ and links $E$\\
$U$ 	& the set of SD pairs\\
$c_e,p_e,q_v$ 	& capacity, ebit generation and swapping success probabilities \\
$s_i,t_i$ 	& the source and destination of the $i$th SD pair\\
$Z_i$ 	& the set of commodities in the $i$th SD pair\\
$z^i_j$ & the $j$th commodity in the $i$th SD pair\\
$d^i_j,a^i_j,\delta^i_j$ & the demand, arrival time and deadline of the commodity $z^i_j$\\
$P_s, P_c$ & the priority list for SD pairs and commodities\\
$P_c^{i}$ & the set of commodities belong to SD pair $i$\\
$P_c^{i}[{{{{l}}}}]$ & the ${{l}}$ commodities in list $P_c^{i}$ \\
$\kappa$ & the scheduling length\\
$\Theta^i_j,\Delta^i_j $ & the remaining demand and time slots of $z^i_j$\\
$\mathcal{S}_T^\tau$ & the network state after Phase-$\tau$ at time $T$\\
$Z_T$ & the set of active commodities at time $T$\\
$\Pi_T^\tau$ & available ebits numbers after Phase-$\tau$ at time $T$\\
$\mathcal{M}_{m{:}n},\mathcal{D}^{m{:}n}_{k{:}n},\mathcal{R}_{m{:}n}$ &  Input buffer \& output buffer \& receiving buffer\\
\hline
Variables & Description\\
\hline
$\eta_{s_it_i}$ & the entanglement distribution rate of $s_i{:}t_i$\\
$f^{m{:}k}_{m{:}n}$ &  amounts of entangled qubit pairs of ${m{:}k}$ would be distributed to ${m{:}n}$ after swapping.\\
$g_{m{:}n}$ & amounts of ebits would be generated along physical link ${m{:}n}$ divided by the capacity $c_{mn}$ \\
\hline
\end{tabular}

\end{table}

\section{Quantum Network Model}
\label{sec:model}
\noindent
In this section, we present preliminaries and modeling of a quantum network. All notations are summarized in Table \ref{notation}.

\subsection{Quantum Information Basics}
\noindent
We consider the prevalent discrete-variable binary state quantum systems. 
A qubit differs from a classical binary bit represented by either $0$ or $1$, and can be in a superposition of two basis states. 
Let $|0\rangle$ and $|1\rangle$ be the two single-qubit basis states.
Here $| \cdot \rangle$ is called a \emph{ket} in Dirac notation denoting a column vector representing the (pure) state of a qubit.
A qubit is written as $|b\rangle = \alpha |0\rangle + \beta |1\rangle$, with complex numbers $\alpha, \beta$ as \emph{amplitudes} of the two basis states, satisfying $|\alpha|^2 + |\beta|^2 = 1$.
Measuring this qubit yields a classical bit of either $0$ with a probability of $|\alpha|^2$ or  $1$ with probability $|\beta|^2$.

A two-qubit system is described by a superposition of four basis states $|00 \rangle$, $|01 \rangle$, $|10 \rangle$ and $|11 \rangle$.
Let $|b_1b_2\rangle = \alpha_{00} |00\rangle + \alpha_{01}|01\rangle + \alpha_{10}|10\rangle + \alpha_{11}|11\rangle$, such that $|\alpha_{00}|^2 + |\alpha_{01}|^2 + |\alpha_{10}|^2 + |\alpha_{11}|^2 = 1$.
If a simultaneous measurement is taken on these two qubits, it will yield either $00$ with probability $|\alpha_{00}|^2$, $01$ with probability $|\alpha_{01}|^2$, $10$ with probability $|\alpha_{10}|^2$, or $11$ with probability $|\alpha_{11}|^2$.
Now, consider a special class of two-qubit states: the four orthogonal Bell states written as
$|\Phi^{\pm}\rangle = \frac{1}{\sqrt{2}}( |00\rangle \pm |11\rangle )$, and $|\Psi^{\pm}\rangle = \frac{1}{\sqrt{2}}( |01\rangle \pm |10\rangle )$.
Two qubits jointly in a Bell state are considered an EPR pair or Bell pair.
Each Bell pair is a \emph{maximally entangled} state because it is the superposition of only two complementary states out of the four basis states, and the two qubits are perfectly correlated.
Bell pairs form the basis of two-party quantum communications: if Alice and Bob each holds one qubit in a Bell pair, they can use it to exchange quantum information by local operations and classical communication.
An entangled qubit pair is called  an \emph{ebit}.

\subsection{Entanglement Generation and Swapping}
\noindent
Entanglements can be distributed over long distances via the generation and swapping of entangled photons.

\textbf{Entanglement generation} is the process of generating an entangled pair of photons that are separated by a physical distance, and each photon in the pair is sent to one of two end nodes through a quantum-capable channel, such as an optical fiber. 
The most common entanglement generation operation is \emph{spontaneous parametric down-conversion (SPDC)}, where a pump laser is shot at a nonlinear crystal and probabilistically generates entangled photon pairs at a high rate.
However, entangled photons will get lost during transmission because of channel attenuation or other environmental factors, and the success probability of entanglement generation will decrease exponentially as distance increases~\cite{pirandola2009direct}. 
For instance, with a typical single-mode fiber loss of $0.2$dB/km~\cite{pant2019routing}, a 10000-pair-per-second entanglement source would only be able to successfully distribute 1 pair per second on average over 200km of distance.
We consider ebits generated and distributed along a physical channel as \emph{elementary ebits}.

\begin{figure}[t]
\centering
\includegraphics[width=0.48\textwidth]{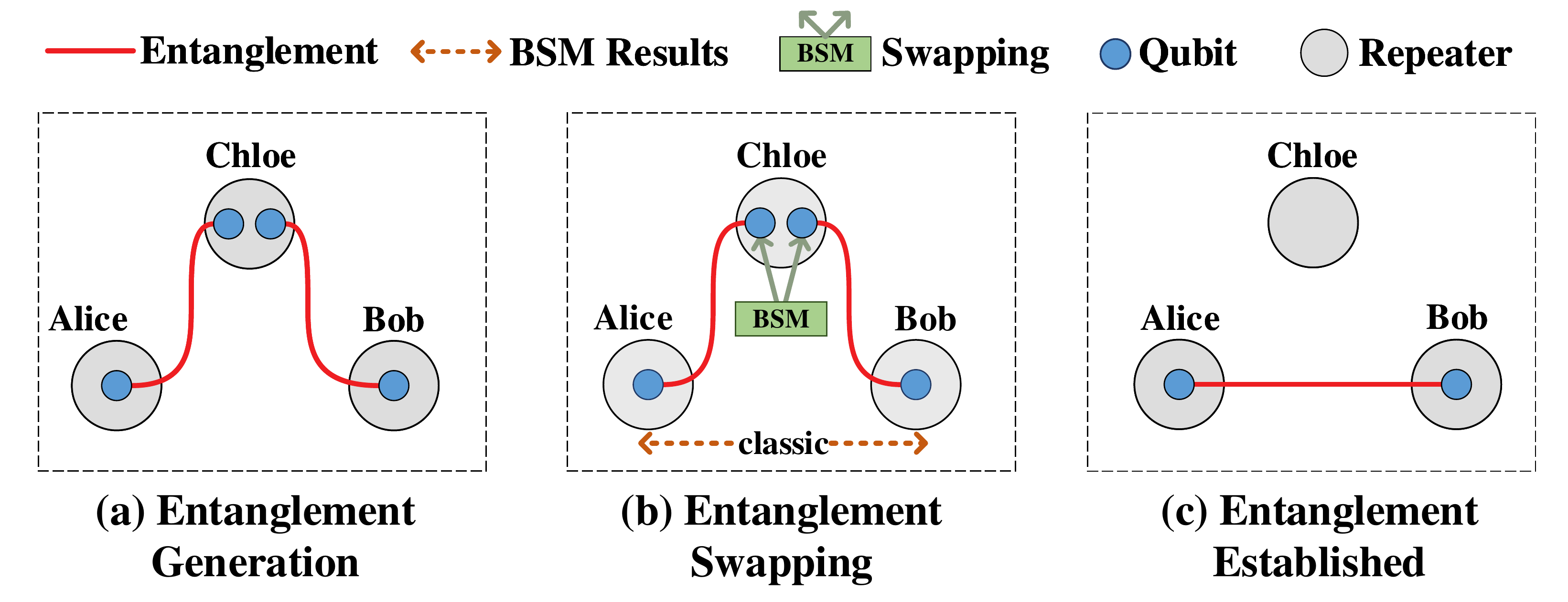}
\caption{\smallfont Entanglement Generation and Swapping}
\label{fig:quantum_process}
\end{figure}

\textbf{Entanglement swapping} is a key process in quantum repeaters to compensate for photon losses in long-distance entanglement generation. 
In Fig.~\ref{fig:quantum_process}, assume Alice and Bob both perform entanglement generation with Chloe, resulting in one ebit between Alice and Chloe and another between Bob and Chloe.
In order to get an end-to-end ebit between Alice and Bob, Chloe will first entangle the two local qubits, and then perform a Bell State Measurement (BSM) to decide which of the four Bell states her local qubits are in. 
She then sends the result to either Alice or Bob, who applies a unitary operation to the local qubit to turn Alice's and Bob's qubits into a Bell pair \emph{without physical interaction}.
Note that because of BSM, this process destroys the two qubits at Chloe regardless of whether the swapping succeeds.
If there are multiple repeaters between Alice and Bob, this process can be repeated recursively until an end-to-end ebit is formed.
The established long-distance end-to-end ebit can then be used for quantum communication without involving any intermediate nodes.

\subsection{Quantum Network Model}
\noindent
We consider a quantum network consisting of quantum repeaters interconnected by lossy links.
The physical topology is denoted by an undirected graph $G = (V, E)$, where node $v \in V$ denotes a quantum repeater and link $e \in E$ is a physical link.
For conciseness, we use $m{:}n$ to interchangeably denote an \emph{unordered} node pair $\{ m, n \}$ for $m, n \in V$; note that $m{:}n = n{:}m$ since they are unordered.
Each node $v \in V$ is associated with swapping success probability $q_v \in (0,1]$.
Each link $e \in E $ consists of $c_e$ quantum channels where $c_e$ is called its \emph{capacity}, and success probability $p_e \in (0,1]$ for entanglement generation along each channel.

We consider a time-slotted quantum network with discrete time $\mathbf{T} = \{ 1, 2, 3, \dots \}$ following existing work~\cite{shi2020concurrent, zhao2021redundant}.
Each time slot $T \in \mathbf{T}$ corresponds to the possible generation and distribution of one ebit along a quantum channel, as well as the completion of one round of swapping at all nodes.
Each time slot is divided into three phases described below:
\begin{enumerate}
\item For any edge $m{:}n\in E$, node $m$ and node $n$ attempt to generate an ebit along each quantum channel with the success probability of $p_{mn}$;
\item For node pair $m{:}n$ and intermediate node $k$ where ebits are available between both $m{:}k$ and $k{:}n$, node $k$ can attempt to perform entanglement swapping with success probability of $q_k$ to establish ebits between $m$ and $n$ by using pairs of ebits between $m{:}k$ and $k{:}n$.
\item 
For SD pair $s_i{:}t_i \in U$, established ebits are distributed to each commodity to be defined in \S\ref{sec:commodity_demand} for quantum communications.
\end{enumerate}

To optimize network performance, we assume a central network controller oversees the entanglement establishment process across the entire network. 
The controller communicates with nodes via classical communication, collects information from network nodes and SD pairs, and makes global entanglement scheduling and routing decisions.

\subsection{Commodities and Demands}
\label{sec:commodity_demand}
\noindent
Consider a set of SD pairs in a quantum network requesting end-to-end ebits for communication.
Following conventional networking terminology, each request is a \emph{commodity}, denoted by $z^i_j \in Z_i
$, where $Z_i$ is the set of all commodities belonging to SD pair $i$.
Each commodity is described by a tuple, $z^i_j = (d^i_j, a^i_j, \delta^i_j)$, where $d^i_j$ denotes the number of requested end-to-end ebits (the demand), $a^i_j$ denotes the arrival time, and $\delta^i_j$ denotes the deadline for finishing the demand.
If a commodity does not have a deadline but wants to be completed as quickly as possible, we define $\delta^i_j = \infty$.
We assume commodities arrive over time, and the network controller has no knowledge of future commodities before they arrive.
Given the time-slotted model, the network state evolves over time.
For ease of description, let $\mathcal{S}_T^\tau$ denote the network state after the Phase-$\tau$ at time slot $T$, for $\tau \! \! \!\in \! \! \! \{0, 1, 2\}$; $\mathcal{S}_T^0$ denotes the state at the beginning of time slot $T$.
Each $\mathcal{S}_T^\tau \! \! \! = \! \! \! (G, Z_T, \Pi_T^\tau)$, where $Z_T  \! \! =  \! \! \bigcup_i Z^i_T$ denotes the set of commodities that are \textbf{active} at time $T$, \kIE, they arrived on or before time $T$ with unfinished demands and unexpired deadlines at time $T$; $\Pi_T^\tau  \!: \!  V \times V  \! \!\rightarrow \mathbb{Z}^*$ denotes the \emph{number of ebits} available between arbitrary node pair $m{:}n$ after Phase-$\tau$ at time slot $T$.
We use $\Theta^i_j$ to denote the unfinished remaining demand of $z^i_j \in Z^i_T$ at time $T$.
The value of $\Pi_T^\tau$ depends on the ebits successfully generated after each Phase-1, the ebits successfully swapped after each Phase-2, and the network model (whether buffers exist; see \S\ref{sec:esdi}).
The goal of the network controller is to meet as many deadlines of commodities as possible, and complete commodities without deadlines as fast as possible, by making real-time decisions on: 1) how to generate ebits over physical links, 2) how to swap ebits at intermediate repeaters, and 3) how to distribute end-to-end ebits to commodities.
\section{Multi-Commodity Entanglement Scheduling and Distribution: A General Framework}
\label{sec:basic}

\noindent
Consider a commodity arrives with a demand and/or a deadline.
The goal of entanglement scheduling and distribution is to deliver as many end-to-end ebits as possible so that the commodity can finish as quickly as possible and hopefully within its deadline.
However, as the overall network entanglement generation rate is bounded by the capacities and further discounted by probabilities of generation and swapping, the primary challenge would be resource contention between multiple commodities with different demands and deadlines.
In the worst case, a fair sharing network may simply lead to all commodities missing deadlines, while some may have succeeded if prioritization between commodities is applied.

We propose \emph{network-wide entanglement scheduling}, combined with \emph{entanglement distribution}, to deal with network resource contention.
Below, we first motivate entanglement scheduling via prioritization, and then define a general framework for optimizing entanglement scheduling and distribution with different scheduling (prioritization) disciplines.
\begin{figure}[t]
\centering
\includegraphics[width=0.48\textwidth]{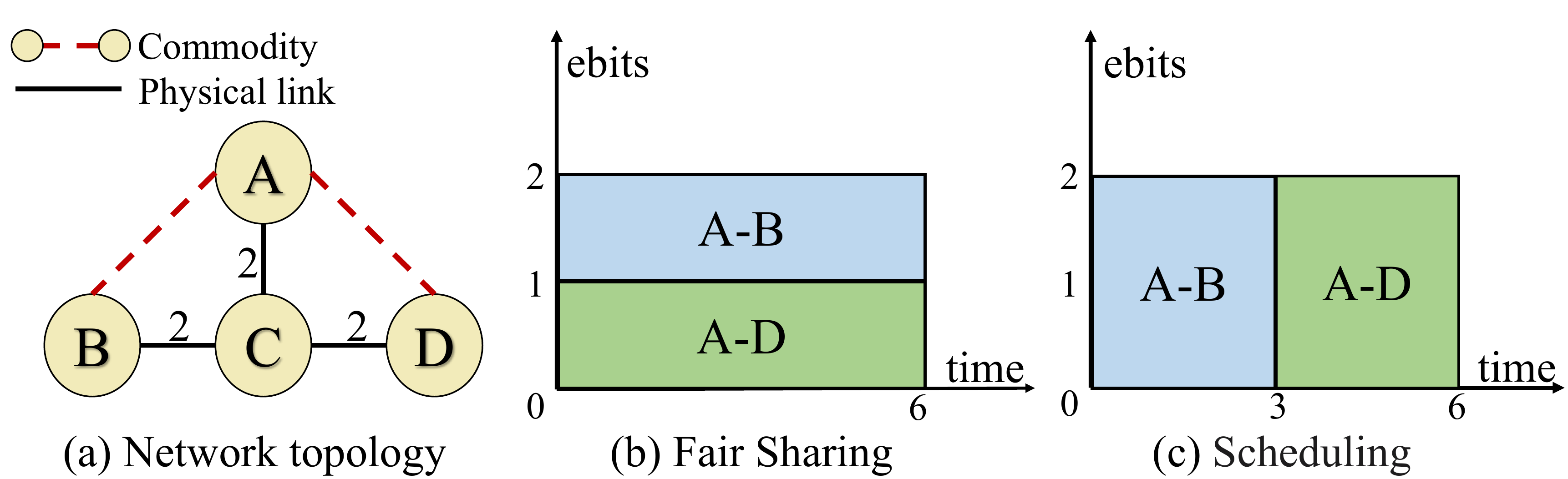}
\caption{\smallfont Motivating example. (a) Network topology with two commodities A-B and A-D and capacity of each link is 2; (b) fair sharing; (c) scheduling with A-B prioritized over A-D.}
\label{fig:scheduling}
\end{figure}

\subsection{A Motivating Example}

\noindent
As shown in Fig.~\ref{fig:scheduling}, consider A, B and D all connected to a quantum repeater C each with a capacity of $2$, and two commodities A-B and A-D arrive at the same time.
For simplicity, assume entanglement generation and swapping are deterministic, \kIE, all success probabilities are $1$.
Both commodities have the same demands of $6$ end-to-end ebits.
In a fair-sharing quantum network, entanglements generated at link A-C are equally shared for swapping to generate A-B and A-D ebits respectively, each at a rate of $1$ ebit per time slot.
Both commodities finish in $6$ time slots, with an \emph{average completion time} of $6$.
Now assume scheduling (prioritization) is employed, where A-B is given priority over A-D.
By utilizing all capacities on links A-C and B-C, commodity A-B can finish within $3$ time slots by generating $2$ end-to-end ebits per time slot, while commodity A-D's completion time remains the same.
The \emph{average completion time} is $\frac{3+6}{2} = 4.5$, yielding a $25\%$ improvement.
If commodity A-B has a deadline of $4$ time slots and commodity A-D has a deadline of $6$ time slots, only A-D can meet its deadline under fair-sharing, while both commodities can meet their deadlines under prioritization, yielding an $100\%$ improvement of \emph{deadline satisfaction ratio}.

The example shows the significance of scheduling in serving time-sensitive communication demands.
Such demands widely exist in near-term quantum applications, due to short coherence time in quantum computers, fast changing object environments in quantum sensing, bursty QKD demands, etc.
We also highlight that the same motivation has been observed and utilized in classical communication such as data center networks, where scheduling of traffic flows can significantly reduce end-to-end latency and improve user experience~\cite{DCTCP, yu2016non}.
The unique characteristics of quantum networking, however, have posed significant challenges in direct application of classical flow scheduling disciplines to quantum networks.

\subsection{Problem Statement}

\noindent
Considering the \emph{probabilistic nature} of quantum operations, in this subsection, we formally define the \emph{entanglement scheduling and distribution (ESDI)} problem.

\begin{definition}
\label{def:sch}
Given a quantum network $G$ and commodities $Z = \bigcup_i Z_i$, a solution to the \textbf{entanglement scheduling and distribution (ESDI)} problem consists of three algorithms, $(\mathcal{A}_{\sf gen}, \mathcal{A}_{\sf swap}, \mathcal{A}_{\sf dis})$ to perform the following tasks respectively:
\begin{itemize}
    \item $\mathcal{A}_{\sf gen}(\mathcal{S}_T^0)$: In Phase-$1$ at time $T$, decide the number of ebits to attempt along physical link $e \in E$;
    \item $\mathcal{A}_{\sf swap}(\mathcal{S}_T^1)$: In Phase-$2$ at time $T$, given the number of ebits between node pairs $m{:}k$ and $k{:}n$ respectively, decide how many ebit pairs are used to swap for node pair $m{:}n$, for $\forall m, k, n \in V$; and
    \item $\mathcal{A}_{\sf dis}(\mathcal{S}_T^2)$: In Phase-$3$ at time $T$, given the number of ebits between each SD pair $s{:}t \in U$, decide how many ebits are distributed to each commodity $z^i_j \in Z_i$.
\end{itemize}
\end{definition}

We note that the above defined problem incorporates all existing entanglement routing or distribution methods as solutions~\cite{Dai2020b, shi2020concurrent, zhao2021redundant, zhao2022e2e, zeng2022multi}, thus enabling fair comparison between existing and new solutions.
Meanwhile, we emphasize that each algorithm in $\{ \mathcal{A}_{\sf gen}, \mathcal{A}_{\sf swap}, \mathcal{A}_{\sf dis} \}$ can be either \emph{deterministic} or \emph{probabilistic}.
Despite this, the inputs to $\mathcal{A}_{\sf swap}$ and $\mathcal{A}_{\sf dis}$ are \emph{always probabilistic} regardless of previous phases' outputs, due to the probabilistic nature of the actual quantum operations (generation and swapping) in the first and second phases.
This probabilistic nature makes the ESDI problem intrinsically challenging.

\subsection{Multi-Commodity Remote Entanglement Distribution}

\noindent
While existing work has tried to design specific algorithms for specific objectives (such as maximizing total entanglement distribution rate (EDR) or ensuring fairness), our first goal is to design a general optimization framework for ESDI that can incorporate flexible objectives and constraints.
In the following, we first extend the single-commodity Optimal Remote Entanglement Distribution (ORED) formulation in~\cite{Dai2020b} to a multi-commodity formulation (MRED), serving as the \emph{backbone} of our optimization framework.

Define variables $\mathcal{F} = \{ f^{m{:}k}_{m{:}n} \ge 0 \,|\, m, k, n \in V \} \cup \{ g_{m{:}n} \in [0, 1] \,|\, (m{:}n) \in E \}$. 
Here $f^{m{:}k}_{m{:}n}$ represents the number of ebits between $m{:}k$ that would be contributed to swapping with ebits between $k{:}n$ at node $k$;
$g_{m{:}n}$ represents the number of ebits that would be attempted to be generated along physical link $m{:}n \in E$, divided by the link capacity $c_{mn}$.
Then, the \textbf{MRED} formulation is defined as follows:
\begin{subequations}
\label{fml:0}
\begin{align}
\smallfont
    \text{(MRED)} &\quad \text{find }
    \mathcal{F}
    \tag{\ref{fml:0}}  
    \\
    \text{s.t.} &\quad 
    f^{m{:}k}_{m{:}n} = f^{k{:}n}_{m{:}n}, 
    \quad \forall k, m, n \in V;
    \label{mored:eq:1}
    \\
    &\quad 
    I(m{:}n) = \Omega(m{:}n), 
    \; \forall m, n \in V, m{:}n \notin U;
    \label{mored:eq:2}
    \\
    &\quad 
    I(s{:}t) \ge \Omega(s{:}t), 
    \; \forall s{:}t \in U.
    \label{mored:eq:3}
\end{align}
Two auxiliary functions $I(m{:}n)$ and $\Omega(m{:}n)$ are defined as:
\begin{align}
    I(m{:}n) \triangleq 1_{m{:}n} p_{mn} c_{mn} g_{m{:}n} \!+\! 
         \sum_{\mathclap{k \in N \setminus \{m, n\}}}\frac{q_k}{2} \left( f^{m{:}k}_{m{:}n} \!+\! f^{k{:}n}_{m{:}n} \right)\!;
         \label{mored:eq:4}
    \\
    \Omega(m{:}n) \triangleq 
         \sum\nolimits_{{k \in N \setminus \{m, n\}}}  
            \left( f^{m{:}n}_{m{:}k} + f^{m{:}n}_{k{:}n} \right),
            \label{mored:eq:5}
\end{align}
\end{subequations}
where $1_{m{:}n} = 1$ if $m{:}n \in E$ and $0$ otherwise.
Here $I(m{:}n)$ denotes the \emph{input} (established) ebits between $m{:}n$, and $\Omega(m{:}n)$ denotes the \emph{output} (consumed) $m{:}n$-ebits for swapping to generate ebits between all other node pairs.

\noindent\textbf{Explanation:} 
Program~\eqref{fml:0} defines a feasibility problem: finding appropriate variables $\{ g_{m{:}n} \}$ (entanglement generation) and $\{ f^{m{:}k}_{m{:}n} \}$ (entanglement swapping), such that: 1) every swapping between $m{:}k$ and $k{:}n$ at node $k$ consumes an equal number of $m{:}k$- and $k{:}n$-ebits as in Constraint~\eqref{mored:eq:1}; 2) for non-SD pairs, all acquired ebits will be used for further swapping as in Constraint~\eqref{mored:eq:2}; 3) for any SD pair $s{:}t$, the input ebits should be no less than the output as in Constraint~\eqref{mored:eq:3}, with the difference corresponding to \emph{end-to-end} ebits that are kept to be used between $s{:}t$ themselves.
Taking a closer look at $I(m{:}n)$, it consists of both ebits generated directly along link $m{:}n$ (the first term), and ebits obtained by swapping at all intermediate nodes $k$ (the second term), both discounted by the corresponding success probabilities $p_{mn}$ and $q_k$ respectively.
The output $\Omega(m{:}n)$ consists of all $m{:}n$-ebits contributed to swapping to generate either $m{:}k$ ebits with $n$ as an intermediate node, or $k{:}n$-ebits with $m$ as an intermediate node.

The MRED formulation itself does not consider online decision making as implicitly included in the ESDI problem definition, since there is no input or decision variable with respect to the time $T$ (or any time-related input).
It nevertheless constitutes a major building block in our framework design due to its ability to incorporate various optimization objectives and constraints.
For example, define $\eta_{st}$ as the total \emph{entanglement generation rate (EDR)} of all commodities between SD pair $s{:}t \in U$, and $\eta \triangleq \sum_{s{:}t \in U} \eta_{st}$ the total EDR of all SD pairs.
The following theorem generalizes Theorem~1 in~\cite{Dai2020b} (with proof omitted due to page limit) showing optimality of the MRED formulation \emph{w.r.t.} the total EDR of all commodities:
\begin{theorem}
\label{th:opt}
The optimal total EDR $\eta^*$ is upper bounded by $\max_{\mathcal{F}} \{ \sum_{s{:}t} (I(s{:}t) - \Omega(s{:}t)) \,|\, \mathcal{F} \text{ is feasible to~\eqref{fml:0}} \}$, and there exists a stationary ESDI protocol with expected total EDR equal to $\eta^*$.
\end{theorem}

\subsection{A General Framework for ESDI}
\label{sec:general_framework}

\noindent
We design a general ESDI framework based on the MRED formulation and the idea of \emph{prioritization} as shown in Algorithm~\ref{a:0}.
Based on the set of commodities and their remaining demands, our framework adjusts the prioritization between SD pairs, by dynamically adding or removing objective functions and constraints to the MRED formulation in~\eqref{fml:0}.
It further balances between \emph{scheduling} (prioritizing certain SD pairs) and \emph{work conservation} (maximizing network EDR).
The solution to the \emph{prioritized} MRED is then executed by a probabilistic protocol to implement the prioritization quickly.

\begin{algorithm}
\caption{\mbox{ESDI General Framework}}
\label{a:0}
$\mathcal{F} \leftarrow \perp$\;
\For{all time slot $T \in \mathbf{T}$ }{
    \If{$\mathcal{F} = \perp$ or priorities may change}{
        \mbox{Adjust objectives and constraints in Program~\eqref{fml:0}};\\[0em]
        Solve adjusted~\eqref{fml:0} to update optimal $\mathcal{F}$\;
    }
    Execute $\mathcal{F}$;
}
\end{algorithm}

In general, the framework keeps track of the set of active commodities and priorities among them.
When initializing, or when priorities among active commodities change, the backbone Program~\eqref{fml:0} will be adjusted with objective functions and/or constraints reflecting the latest priorities.
It then gets solved to derive the optimal solution $\mathcal{F}$, which will be executed over time until priorities change again.
The priority changes are decided by a \textbf{scheduling algorithm} and could happen in a number of cases such as changes in active commodities and demands.
The solution $\mathcal{F}$ is instead executed by a \textbf{distribution protocol} consisting of $\{ \mathcal{A}_{\sf gen}, \mathcal{A}_{\sf swap}, \mathcal{A}_{\sf dis} \}$ as defined in Definition~\ref{def:sch}.
In the following, we will focus on the design of these elements, including two {scheduling algorithms} for deadline-agnostic (\S\ref{sec:sjf}) and deadline-aware (\S\ref{sec:edf}) commodities respectively, and a generic distribution protocol that can be tailored to both types of commodities (\S\ref{sec:esdi}).
\section{ESDI without Deadline Constraints}
\label{sec:sjf}
\noindent
In this section, we consider ESDI for commodities without deadlines.
An example is QKD between end points~\cite{liao2017satellite,peev2009secoqc}, whose focus is to accumulate a sufficient number of classical key bits obtained from entanglements as soon as possible.
The main goal of scheduling is to minimize the average completion time of all commodities in a quantum network.
\subsection{Motivation: Shortest Job First Scheduling}

\noindent
\emph{Shortest job first (SJF)} and its variant, \emph{shortest remaining time first (SRTF)}, are optimal scheduling policies for average task completion time in classical real-time task scheduling. 
SJF allocates the idle processing unit to the task with the shortest completion time among the remaining tasks, while SRTF allows a newly arriving task to \emph{preempt} the currently processing task. 
Due to their strong performance and simple implementation, SJF and SRTF are widely used in network traffic scheduling~\cite{PIAS,pdq}.
Nevertheless, extending these policies to ESDI is very challenging.
First, scheduling over a network of repeaters is more challenging than scheduling on a single machine and is generally NP-hard~\cite{han2015rush}.
In this case, exclusively scheduling one commodity at a time is clearly inefficient since the network may simultaneously support multiple commodities without resource contention.
Second, the probabilistic nature of quantum operations makes it impossible to obtain the exact completion time of each commodity.
Our goal is to design an algorithm that \emph{simulates} the behavior of an SJF scheduler in the quantum network scenario, while balancing between \textbf{strict scheduling} (one commodity at a time) and \textbf{work conservation} (achieving high utilization of network resources and throughput). %
The design of our algorithm includes two components: 1) an MRED formulation with strict priorities, and 2) an SJF-based prioritization algorithm.

\subsection{MRED with Strict Priorities}

\noindent
Consider a \emph{priority list of SD pairs $P_s$} given as input, which consists of $\kappa$ SD pairs sorted from high to low priority.
Let $\eta_i = \eta_{s_it_i}$ be the expected EDR between SD pair $s_i{:}t_i \in P_s$ with index $i$.
The MRED with strict priorities (MRED-SP) is formulated by enforcing strict priorities among commodities:
\begin{subequations}
\label{fml:mored_sjf}
\begin{align}
\text{(MRED-SP)}
&\quad
\max \eta_{1},\; \max \eta_{2},\; \dots,\; \max \eta_{\kappa},\;
\nonumber
\\
&\quad \max \sum\nolimits_{{s{:}t \in U} }\eta_{st} 
\tag{\ref{fml:mored_sjf}} 
\\
\text{s.t.}
&\quad
\eta_{st} = I(s{:}t) - \Omega(s{:}t),
\quad \forall s{:}t \in U;
\label{mored_sjf:eq:1}
\\
&\quad
\nonumber
\text{Constraints~\eqref{mored:eq:1}--\eqref{mored:eq:5}}.
\end{align}
\end{subequations}

\noindent\textbf{Explanation:} 
Program~\eqref{fml:mored_sjf} is a \emph{multi-objective optimization} problem with up to $\kappa + 1$ objectives.
The first $\kappa$ objectives enforce strict priorities among SD pairs, \emph{i.e.}, the program will first optimize for $\eta_1$, then optimize for $\eta_2$ while keeping optimality of $\eta_1$, then optimize for $\eta_3$ while keeping optimality of both $\eta_1$ and $\eta_2$, so on and so forth.
The last objective, which optimizes for the total EDR of all SD pairs, is added to achieve work conservation, \kIE{}, utilizing the remaining resources unused by the $\kappa$ prioritized SD pairs to maximize the overall throughput of the network.
With Constraint~\eqref{mored_sjf:eq:1} providing a definition of each $\eta_{st}$, Program~\eqref{fml:mored_sjf} is an optimization version of Program~\eqref{fml:0} enforcing constraints~\eqref{mored:eq:1}--\eqref{mored:eq:5}.

\noindent\textbf{Remark:}
We note that while the list $P_s$ may also be defined upon commodities instead of SD pairs, it is not meaningful in the formulation.
Consider two commodities belonging to the same SD pair $s{:}t$.
Incorporating both in the priority list results in adding the objective $\eta_{st}$ twice to Program~\eqref{fml:mored_sjf}, which has no impact than just adding one for the higher-priority commodity, but will increase solving time of the multi-objective program.%

\subsection{ESDI-O: SJF-based Priority Scheduling}
\noindent 

\noindent
The key in utilizing Program~\eqref{fml:mored_sjf} for ESDI is to form the priority list $P_s$ of SD pairs.
Following the intuition of SJF, $P_s$ should reflect how fast each SD pair will be able to finish its commodity, \kIE{}, the \emph{expected completion time (ECT)} of the commodity, with the lowest demand.
This is however challenging because of the difficulty in modeling the exact probability distributions governing the entanglement generation and swapping processes across the network.
Nevertheless, as our primary goal is to define priorities instead of obtaining the accurate ECTs, we may use an approximation---a lower bound of the ECT.
Specifically, consider a commodity $z^i_j \in Z_i$ receives an expected EDR of $\eta$, and define a random variable $\eta(T)$ as the actual number of ebits generated and distributed to $z^i_j$ in every time slot $T$.
We have $ECT^i_j\!\!\! =\!\!\! \mathbb{E}\left[ \min\left\{ T' \,\left|\, \sum_{T=1}^{T'} \eta(T) \ge d^i_j \right\} \right. \right] 
\!\!\!\approx \!\!\!\mathbb{E}\left[ \min\left\{ T' \,|\, \mathbb{E}[\eta] \cdot T' \ge d^i_j \right\} \right]\!\!\! = \!\!\!\mathbb{E}[ d^i_j / \mathbb{E}[\eta] ]
$.
By Jensen's inequality, we then have $ECT^i_j \ge d^i_j / \mathbb{E}[\eta]$.

To rank SD pairs by priorities, we use $d^i_j / \mathbb{E}[\eta]$ as an approximate ECT for each commodity.
Here $\mathbb{E}[\eta]$ is computed by running MRED with only one SD pair in the quantum network, \kIE{}, exclusively allocating all network resources to a single SD pair.
SD pairs are then ranked by the approximate ECT of the commodity with the \emph{smallest demand} for each pair.
After ranking, a multi-commodity (multi-objective) MRED-SP formulation is solved to obtain the eventual ESDI solution.

\begin{algorithm}
\caption{\mbox{ESDI-O for Commodities without Deadlines}}
\label{a:3}
\KwIn{Network $G = (V,E)$, commodities $Z$, SD pairs $U$, scheduling length $\kappa$}
$\mathcal{F} \leftarrow \emptyset$, $Z_T \leftarrow \emptyset$ for $\forall T$, $\eta_{st} \leftarrow \perp$ for $\forall s{:}t \in U$\;\label{alg3:init}
\For{SD pair $s{:}t \in U$}{
    $\eta_{st} \leftarrow \max_{\mathcal{F}} \{ \eta_{st} \,|\, \text{Program~\eqref{fml:0}} \}$\;
}\label{alg3:eta}
\For{$T = 1, 2, \dots$}{
    Add arriving commodities to $Z_T$\;
    \If{active commodity list $Z_T$ has changed}{
        $P_s$ $\leftarrow$ sort SD pairs by $\min_j\{d^i_j\} / \eta_{s_it_i}$\;\label{alg3:sort}
        Keep first $\kappa$ SD pairs in $P_s$ and drop the rest\;
        $\mathcal{F} \leftarrow$ solve Program~\eqref{fml:mored_sjf} with $P_s$\;\label{alg3:solve}
    }
    Execute $\mathcal{F}$ by calling $(\mathcal{A}_{\sf gen}, \mathcal{A}_{\sf swap}, \mathcal{A}_{\sf dis})$\;\label{alg3:exe}
        $Z_{T+1} \leftarrow Z_T \setminus \{ z^i_j \,|\, d^i_j \text{ is finished} \}$\;\label{alg3:complete}
}
\Return{when all commodities have finished.}
\end{algorithm}

Based on the above intuition,
Algorithm~\ref{a:3} gives the detailed online ESDI-O algorithm to minimize the average completion time of all commodities.
Line~\ref{alg3:init} initializes the solution $\mathcal{F}$, the active commodity list $Z_T$ and the expected EDR $\eta_{st}$ for every SD pair.
In Line~\ref{alg3:eta}, the expected EDR for each SD pair is computed \emph{offline}, by assuming it is the only SD pair in the network.
In the online process, whenever the active commodity list $Z_T$ changes with either commodity arriving at time $T$ or was completed after $T-1$, the solution $\mathcal{F}$ will be updated.
First, in Line~\ref{alg3:sort}, SD pairs will be sorted by \emph{lower bound} of the ECT of each SD pair's commodity with the \emph{lowest demand}, \kIE{}, $\min_j\{d^i_j\} / \eta_{s_it_i}$.
The first $\kappa$ SD pairs are then entered into Program~\eqref{fml:mored_sjf} in the priority list $P_s$, and the solution $\mathcal{F}$ will be updated after solving Program~\eqref{fml:mored_sjf} in Line~\ref{alg3:solve}.
Then, in every time slot, the up-to-date solution $\mathcal{F}$ is executed, by calling $(\mathcal{A}_{\sf gen}, \mathcal{A}_{\sf swap}, \mathcal{A}_{\sf dis})$ which we detail in \S\ref{sec:protocol}.
Finally, completed commodities will be removed from the active commodity list $Z_{T+1}$ in Line~\ref{alg3:complete}.

\noindent\textbf{Remark:}
In addition to following the general intuition of SJF that we outlined before, Algorithm~\ref{a:3} also contains several practical design elements to improve its performance in practice.
First, the solution $\mathcal{F}$ is only updated when the commodity list changes.
Second, when scheduling with different sets of commodities, we seek to \emph{preserve the relative priorities between commodities}, by sorting only based on the original demand $d^i_j$ instead of the remaining demand of each commodity.
While these may seem counter-intuitive from the scheduling perspective, we find that they actually help improve the performance by achieving better \emph{work conservation}.
Specifically, because the network generates ebits probabilistically between each pair of swapping node pairs $m{:}k$ and $k{:}n$, almost in any time slot $T$ there is one side with more generated ebits than the other side, leading to intermediate ebits not being utilized in time $T$.
These intermediate ebits would be wasted if the underlying swapping decisions have changed, wasting resources and degrading throughput.
By updating the solution $\mathcal{F}$ only when active commodities change, and preserving relative priorities among commodities, the algorithm can minimize the number of times that the underlying swapping decisions change, hence reducing wastage and accelerating the completion of commodities.
\section{ESDI With Deadline Constraints}
\label{sec:edf}
\noindent
In this section, we consider ESDI for commodities with deadlines. 
In some quantum applications like DQC, decoherence is a critical challenge, where the quantum information stored in qubits gradually decoheres over time (even with quantum error correction)~\cite{cacciapuoti2019quantum}.
It is thus crucial to finish transmitting the information before irreversible errores happen.%

\subsection{Motivation: Earliest Deadline First Scheduling}
\noindent
When scheduling tasks with deadlines, \emph{Earliest Deadline First (EDF)} is a provably optimal \emph{preemptive} scheduling policy in classical real-time scheduling~\cite{pdq}. 
But similar to SJF, it cannot be directly extended due to the quantum network characteristics, more specifically, difficulty in accurately estimating the ECT and the multi-resource contention among multiple commodities.
Further, the previous formulation in~\eqref{fml:mored_sjf} is no longer suitable because of deadlines, since it would prioritize one commodity strictly over another, neglecting cases where a set of deadlines may mostly or all be satisfied when the commodities jointly share the network resources.

This section develops an EDF-inspired algorithm following the same structure as in the last section, including: 1) an MRED formulation enforcing deadline-aware priorities, and 2) an EDF-based prioritization algorithm.

\subsection{MRED with Deadline Constraints}
\noindent
Here, instead of giving a list of SD pairs with strict priorities among them, we are given a \emph{priority list of commodities} $P_c$, such that we seek to make sure \emph{all} commodities in $P_c$ can be completed by their deadlines \emph{on expectation}.
Each commodity $z^i_j \in P_c$ has a \emph{remaining demand} $\Theta^i_j$ denoting the number of end-to-end ebits yet to be distributed, and $\Delta^i_j$ remaining time slots until the deadline of the commodity.
Let $U_c$ be the set of SD pairs that commodities in $P_c$ belong to, and let $P_c^i \subseteq P_c$ be the set of commodities belonging to $i \in U_c$.
The commodities in each $P_c^i$ are sorted in \textbf{non-decreasing order of their remaining time slots $\Delta^i_j$}.
We further facilitate notation by defining $P_c^{i}[{{{{l}}}}]$ as the \emph{first ${{{{l}}}}$ commodities in list $P_c^{i}$}.

Note that SD pairs in $U_c$ are not prioritized over each other, \emph{i.e.}, all SD pairs are treated the same; but priorities are defined among commodities belonging to each SD pair in the priority list $P^i_c$.
Let $\eta_{s_it_i}$ be the expected EDR between SD pair $s_i{:}t_i \in U$ as previously defined.
The MRED with deadline constraints (MRED-DC) is formulated as followed:

\begin{subequations}
\label{fml:mored_edf}
\begin{align}
\text{(MRED-DC)}
&\quad \max \;\; \sum\nolimits_{s{:}t \in U} \eta_{st}\tag{\ref{fml:mored_edf}} 
\\
\text{s.t.}
&\quad
\begin{aligned}
\eta_{s_it_i}\Delta^i_j  &\geq  \sum\nolimits_{z^i_{j}\in P^i_c[{{{{l}}}}]}{\Theta^i_{j}}, \\
&\forall s_i{:}t_i \in U_c, {{{{l}}}} = 1, 2, \dots, |P_c^i|;
\end{aligned}
\label{fml:mored_edf:b}
\\
&\quad
\nonumber
\text{Constraints~\eqref{mored:eq:1}--\eqref{mored:eq:5} and~\eqref{mored_sjf:eq:1}}.
\end{align}
\end{subequations}

\noindent\textbf{Explanation:} 
Different from Program~\eqref{fml:mored_sjf} in which the objectives enforce strict priorities, the objective function in Program~\eqref{fml:mored_edf} is merely to achieve \emph{work conservation}, \kIE{}, maximizing overall network throughput.
Instead, the prioritization is fully enforced through Constraint~\eqref{fml:mored_edf:b}, which specifies \textbf{\emph{requirements}} on the expected EDR of each SD pair in the priority set $U_c$.
Consider an SD pair $i$ with only one commodity $z^i_j \in P_c^i$.
Constraint~\eqref{fml:mored_edf:b} basically specifies that the expected number of ebits distributed within the remaining time slots $\Delta^i_j$, must be able to satisfy all the remaining demand $\Theta^i_j$ of the commodity, that is, $\eta_{s_it_i} \Delta^i_j \ge \Theta^i_j$.

The case gets trickier when an SD pair has multiple commodities in $P_c^i$.
In this case, having $\eta_{s_it_i} \Delta^i_j \ge \Theta^i_j$ for each individual $z^i_j \in P_c^i$ is no longer sufficient, since the distributed end-to-end ebits must be shared among commodities.
Nevertheless, a close inspection reveals that with respect to a single SD pair, the contention among commodities precisely replicates the classical single-machine task scheduling problem with deadlines, in which case EDF is proved to be optimal.

Indeed, Constraint~\eqref{fml:mored_edf:b} fully simulates EDF with respect to each SD pair, by always requiring the commodity with the smallest remaining time slot $\Delta^i_j$ to be completed first.
The first commodity $z^i_{j_1} \in P_c^i[1]$ thus satisfies the same condition $\eta_{s_it_i} \Delta^i_{j_1} \ge \Theta^i_{j_1}$.
For the second commodity $z^i_{j_2} \in P_c^i[2] \setminus P_c^i[1]$, it can only start after $z^i_{j_1}$ has finished, and hence the condition for it to finish becomes $\eta_{s_it_i} \Delta^i_{j_2} \ge \Theta^i_{j_1} + \Theta^i_{j_2}$.
The third commodity similarly requires $\eta_{s_it_i} \Delta^i_{j_3} \ge \Theta^i_{j_1} + \Theta^i_{j_2} + \Theta^i_{j_3}$, so on and so forth.
Constraint~\eqref{fml:mored_edf:b} gives the general form.%
\noindent\textbf{Remark:} 
MRED-DC prioritizes commodities via constraints, which is generally stronger than prioritization via objectives.
This is to enforce the deadline requirement of commodities.
Meanwhile, MRED-DC can be more \emph{work conserving} than MRED-SP, in the sense that each prioritized commodity or SD pair does not try to take all the network resources, but instead would only take what is needed to complete before the deadline, leaving more room for network-wide throughput optimization.
Note that despite MRED-DC requiring completion \emph{on expectation}, a prioritized commodity may not be able to finish after all due to statistical inevitability.
But we find such cases relatively rare and do not impact MRED-DC's practical performance with the prioritization algorithm in \S\ref{sec:edf-scheduling}.

On the other hand, one implication of prioritization via constraint is the possibility of an \textbf{infeasible} program, in which it may not be possible for all input commodities to complete on expectation (due to too short remaining deadlines or too much contention).
We address this by the following algorithm.

\subsection{ESDI-E: EDF-based Priority Scheduling}
\label{sec:edf-scheduling}
\noindent 
The essence of utilizing Program \eqref{fml:mored_edf} is to adjust the commodity priority list based on the active commodities dynamically.
The priority list $P_c$ should be able to reflect how many time slots commodities have for completing their demands. 
At time $T$, the remaining time slots for commodity $z^i_j$ is $\Delta^i_j = \delta^i_j - T + 1$.

\begin{algorithm}
\caption{\mbox{ESDI-E for Commodities with Deadlines}}
\label{a:1}
\KwIn{Network $G = (V,E)$, commodities $Z$, SD pairs $U_T$, scheduling length $\kappa$}
$\mathcal{F} \leftarrow \emptyset$, $Z_T \leftarrow \emptyset$ for $\forall T$, $\eta_{st} \leftarrow \perp$ for $\forall s{:}t \in U$\;\label{alg1:init}
\For{$T = 1, 2, \dots$}{
    $Z_T \leftarrow Z_T \setminus \{ z^i_j \,|\, \delta^i_j < T \}$\;\label{alg1:drop_deadline}
    Add arriving commodities to $Z_T$\;
    $\Delta^i_j \leftarrow \delta^i_j - T + 1$ for $\forall z^i_j \in Z_T$\;
    \If{active commodity list $Z_T$ has changed}{\label{alg1:enter}
        $P_c \leftarrow \emptyset$\;
        $\mathcal{F} \leftarrow$ solve Program~\eqref{fml:mored_edf} with $P_c$\;
        \For{$z^i_{j} \in Z_T$ in increasing order of $\Delta^i_j$ }{\label{alg1:sort}
            \lIf{$|P_c| = \kappa$}{\Break}
            $P_c \leftarrow P_c \cup \{ z^i_j \}$\;
            $\mathcal{F} \leftarrow$ solve Program~\eqref{fml:mored_edf} with $P_c$\;
            \lIf{$\mathcal{F}$ is infeasible}{$P_c \leftarrow P_c \setminus \{ z^i_j \}$}
        }

    }
    Execute last feasible $\mathcal{F}$ via $(\mathcal{A}_{\sf gen}, \mathcal{A}_{\sf swap}, \mathcal{A}_{\sf dis})$\;\label{alg1:exe}
    Update remaining demands $\Theta^i_j$ for $z^i_j \in Z_T$\;
        $Z_{T+1} \leftarrow Z_T \setminus \{ z^i_j \,|\, d^i_j \text{ is finished} \}$\;\label{alg1:complete}
}
\Return{when all commodities have finished.}
\end{algorithm}

Algorithm~\ref{a:1} provides the detailed online ESDI-E algorithm to maximize the number of commodities that could be finished before their deadlines.
Line~\ref{alg1:init} initializes the solution $\mathcal{F}$, the active commodity list $Z_T$, and the expected EDR $\eta_{st}$ for every SD pair.
For the online process, any commodity not finished by its deadline will be dropped in Line~\ref{alg1:drop_deadline}.
The solution $\mathcal{F}$ and commodities in the priority list $P_c$ will be updated once new commodities arrive or commodities leave at time slot $T$, starting from Line~\ref{alg1:enter}.
Active commodities will be sorted by the remaining deadlines $\Delta^i_j$ to ensure commodities with more urgent deadlines could be served as soon as possible in Line~\ref{alg1:sort}. 

The priority list $P_c$ is built \emph{incrementally} via sequentially solving Program~\eqref{fml:mored_edf}, with the goal of forming a \emph{maximally feasible} priority list with up to $\kappa$ prioritized commodities.
In each iteration at Line~\ref{alg1:sort}, one commodity is added to $P_c$, and Program~\eqref{fml:mored_edf} is solved to decide if all commodities in $P_c$ can finish by their deadlines on expectation.
If adding a commodity makes Program~\eqref{fml:mored_edf} infeasible, it will be dropped from $P_c$.
The eventual solution $\mathcal{F}$ at time $T$ is the last feasible solution of Program~\eqref{fml:mored_edf}.
The rest for executing the solution $\mathcal{F}$ and updating the active commodity list are the same as in Algorithm~\ref{a:3}.

\noindent\textbf{Remark:} 
For the same reason in ESDI-O (minimizing wasted intermediate ebits), ESDI-E also updates the solution only when the commodity list changes.
Sorting commodities by remaining deadlines naturally preserves the relative priorities among commodities, similar to using the original demands for sorting in ESDI-O.
Meanwhile, ESDI-E may be significantly slower than ESDI-O with a large $\kappa$, due to the need of repeatedly solving Program~\eqref{fml:mored_edf} in each update iteration to ensure feasibility.
For this reason, $\kappa$ should be kept as a small value, which additional benefits work conservation for maximizing network throughput.

\section{\textbf{ESDI}: Entanglement Distribution Design}
\label{sec:esdi}
\noindent
With the solution $\mathcal{F}$ output by either ESDI-O or ESDI-E, we next design algorithms $\{ \mathcal{A}_{\sf gen}, \mathcal{A}_{\sf swap}, \mathcal{A}_{\sf dis} \}$ to implement the solution, targeting at a buffered quantum network scenario.

\subsection{Buffered Quantum Network}
\begin{figure}[t]
\centering
\includegraphics[width=0.48\textwidth]{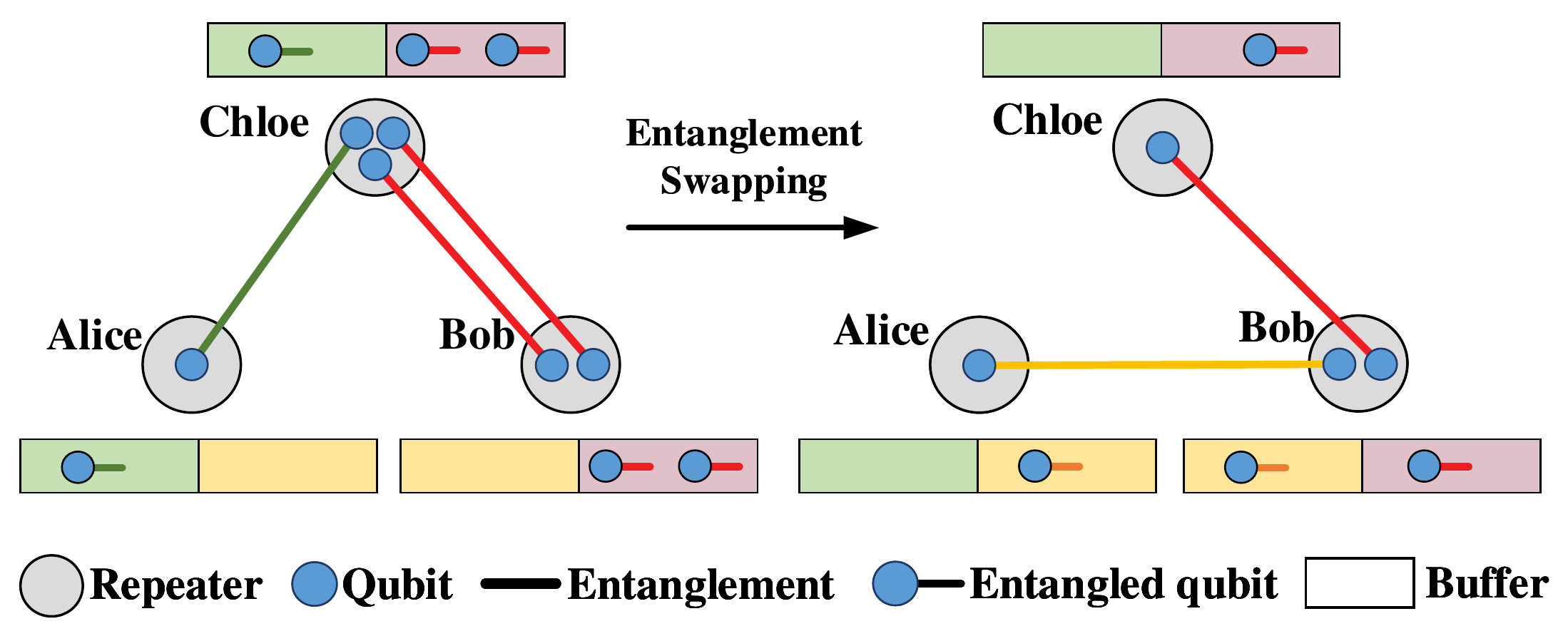}
\caption{Buffered quantum network. Different colors of buffers and entanglements represent half part of an ebit being stored at the corresponding repeater's buffer and entanglement between the corresponding repeaters, respectively.}
\label{fig:buffered_quantum}
\end{figure}
\noindent
In existing work on entanglement routing or distribution, it is commonly assumed that quantum memories are unavailable or have extremely short coherence time~\cite{shi2020concurrent, zhao2021redundant, zhao2022e2e, zeng2022multi}. 
Hence ebits generated for swapping must be either consumed or discarded in one time slot. This severely limits the entanglement generation rate of the quantum network.
Nevertheless, recent advances in quantum memories have demonstrated relatively long storage times of coherent qubits from seconds~\cite{dudin2013light} to over 1 hour~\cite{ma2021a}.
This has motivated recent designs of \emph{buffered quantum networks}, where intermediate repeaters can store ebits for an extended period before entanglement swapping~\cite{Dai2020b,gu2023fendi,Dai2020a}.
Below, we first describe a general model for a buffered quantum network and then present protocols implementing our optimized entanglement scheduling and distribution solutions.
As is shown in Fig.~\ref{fig:buffered_quantum},
We consider establishing ebits between one SD pair Alice and Bob, with the help of repeater Chloe. 
Assume after the entanglement generation, there are one entanglement between Alice and Chloe and two entanglements between Chloe and Bob.
Each ebit would be stored in buffers at both ends, respectively. 
Now Chloe performs entanglement swapping for one ebit pair from each buffer.
If swapping succeeds, an ebit is generated between Alice and Bob while the two qubits in Chloe's buffer are both consumed. 
After getting the swapping result, Alice would move her local qubit from her buffer with Chloe to her buffer with Bob, and Bob would move his local qubit from his buffer with Chloe to his buffer with Alice, which can be done by simply relabeling these qubits. 
The remaining unused ebits between Chloe and Bob would be stored in Chloe's and Bob's buffers to wait for another Alice-Chloe pair for swapping.

\subsection{Entanglement Distribution Design}
\label{sec:protocol}
\noindent 
Given a solution $\mathcal{F}$ output by a central quantum network controller, 
we design a \emph{multi-commodity} extension of the protocol in~\cite{Dai2020a} to achieve the expected entanglement distribution rate. %
For the generation process ($\mathcal{A}_{\mathsf{gen}}$), each link $m{:}n$ will {continuously attempt to generate elementary ebits at rate $c_{mn} \cdot g_{m{:}n}$}.
For any node pair $m{:}n$, we consider set $\mathcal{M}_{m{:}n}$ to store established ebits between node $m$ and node $n$, and set $\mathcal{D}^{m{:}n}_{m{:}k}$ to store $m{:}n$-ebits that will be used to swap into $m{:}k$-ebits.
For an SD pair $s{:}t \in U$, an additional set $\mathcal{R}_{s{:}t}$ is kept to store all end-to-end ebits distributed between source $s$ and destination $t$.
All sets are physically implemented by \emph{quantum buffers} with classical labels at both $m$ and $n$, with $\mathcal{M}_{m{:}n}$ denoting the \emph{\textbf{input buffer}} of the node pair, $\mathcal{D}^{m{:}n}_{m{:}k}$ denoting the \emph{\textbf{output buffer}} for ``next-hop'' node pair $m{:}k$, and $\mathcal{R}_{m{:}n}$ denoting the \emph{\textbf{receiving buffer}} for storing ``completed'' end-to-end ebits to be utilized by upper-layer quantum communication protocols or applications.

The swapping process ($\mathcal{A}_{\mathsf{swap}}$) is a probabilistic process with two steps: 1) moving generated/established ebits from the input buffer to output buffers \emph{probabilistically} (switching), and 2) swapping when corresponding buffers are non-empty (swapping).
Whenever an ebit is added to $\mathcal{M}_{m{:}n}$, the two endpoints will implement \emph{\textbf{opportunistic switching}}, by jointly tossing a random coin and moving the ebit from $\mathcal{M}_{m{:}n}$ to $\mathcal{D}^{m{:}n}_{m{:}k}$ or $\mathcal{R}_{m{:}n}$ with the following probabilities:
\begin{align*}
    \smallfont
    \Pr[\text{move to }\mathcal{D}^{m{:}n}_{m{:}k}] = \frac{f^{m{:}n}_{m{:}k}}{\sum_{k'}{ f^{m{:}n}_{m{:}k'} } + \eta_{mn} }, \\[0.3em]
    \Pr[\text{move to }\mathcal{R}_{m{:}n}] = \frac{\eta_{mn}}{\sum_{k'}{ f^{m{:}n}_{m{:}k'} } + \eta_{mn} }.
\end{align*}
where $\eta_{mn} = 0$ for a non-SD pair $m{:}n \notin U$.
Finally, each node $k$ will check if for any $m{:}n$, there exists 
\begin{enumerate}
    \item $f^{m{:}k}_{m{:}n} = f^{k{:}n}_{m{:}n} > 0$; \\[-0.7em]
    \item $\mathcal{D}^{m{:}k}_{m{:}n} \ne \emptyset$, and $\mathcal{D}^{k{:}n}_{m{:}n} \ne \emptyset$.
\end{enumerate}
For each such case, node $k$ performs swapping between each pair of ebits in $\mathcal{D}^{m{:}k}_{m{:}n}$ and $\mathcal{D}^{k{:}n}_{m{:}n}$ respectively.
Upon success, the ebit will then be added to $\mathcal{M}_{m{:}n}$ by $m$ and $n$.
Compared to the single-commodity protocol in~\cite{Dai2020a}, the additional receiving buffer $\mathcal{R}_{s{:}t}$ is required for implementing MRED, since each SD pair may also be assigned to contribute established ebits for other SD pairs, instead of keeping all ebits to their own use.
All the above can be parallel and asynchronous.  %
Once end-to-end ebits are established and stored in $\mathcal{R}_{s{:}t}$, it further needs to be distributed to commodities of the same SD pair.
This distribution process ($\mathcal{A}_{\mathsf{dis}}$) can be deterministic.
Following our design principles in \S\ref{sec:sjf} and \S\ref{sec:edf}, we apply the following policies:
\begin{enumerate}
    \item For commodities without deadlines, the commodity with the least remaining demand will be served first, followed by the second, so on and so forth.
    \item For commodities with deadlines, the commodity with the earliest deadline would be served, followed by the second, so on and so forth.
\end{enumerate}
Both are due to that commodity scheduling for single SD pair reduces to (deterministic) single-machine classical scheduling, where SJF and EDF are optimal in corresponding scenarios. 
\begin{figure*}[h]
\vspace{-1em}
\subfloat[Impact of the arrival rate ]{\includegraphics[width=0.28\textwidth]{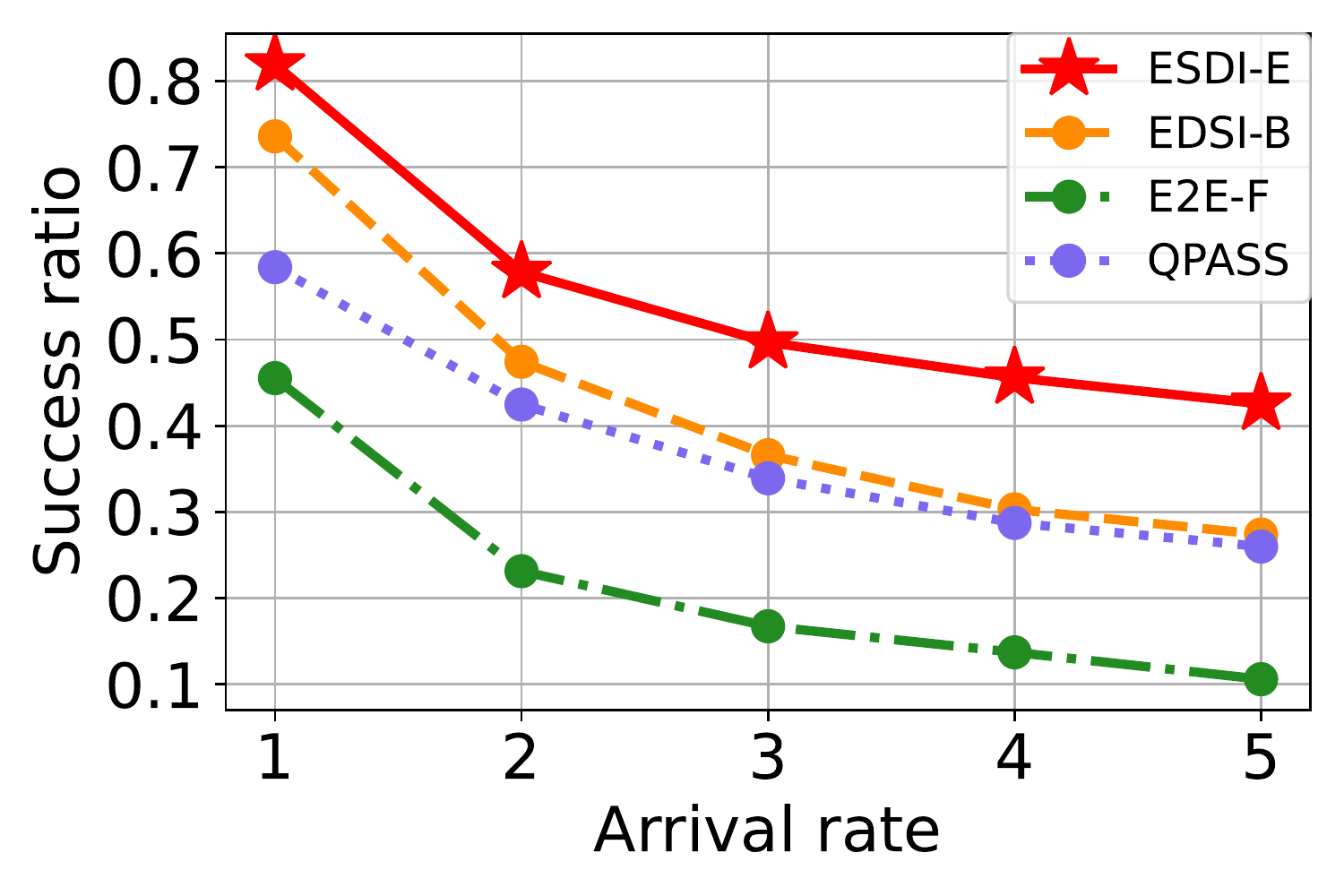}
\label{fig:algs_ddl:success_ratio_pos_ddl}}
\hfil
\subfloat[Impact of the demand]{\includegraphics[width=0.28\textwidth]{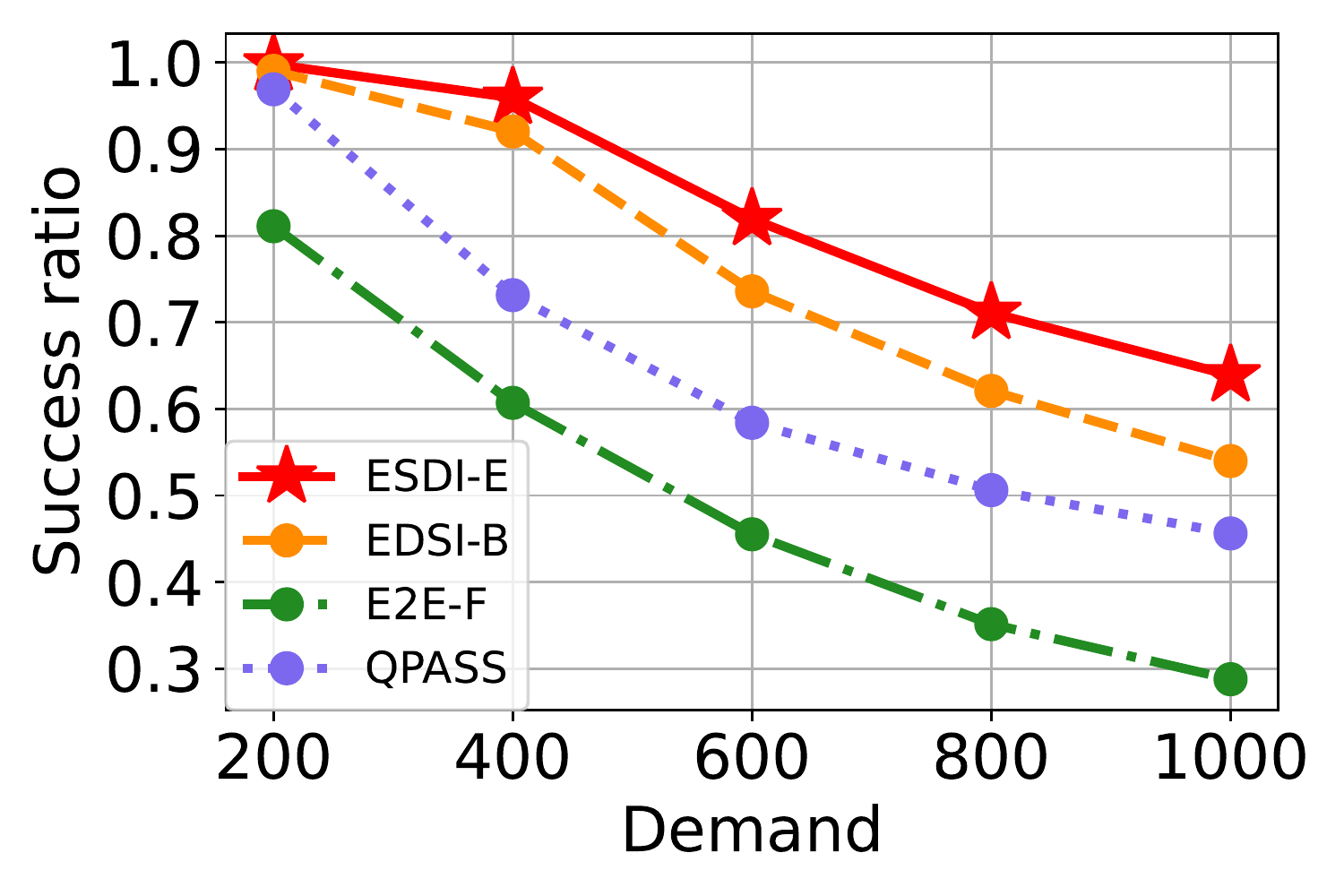}
\label{fig:algs_ddl:success_ratio_demand_ddl}}
\hfil
\subfloat[Impact of the graph size]{\includegraphics[width=0.28\textwidth]{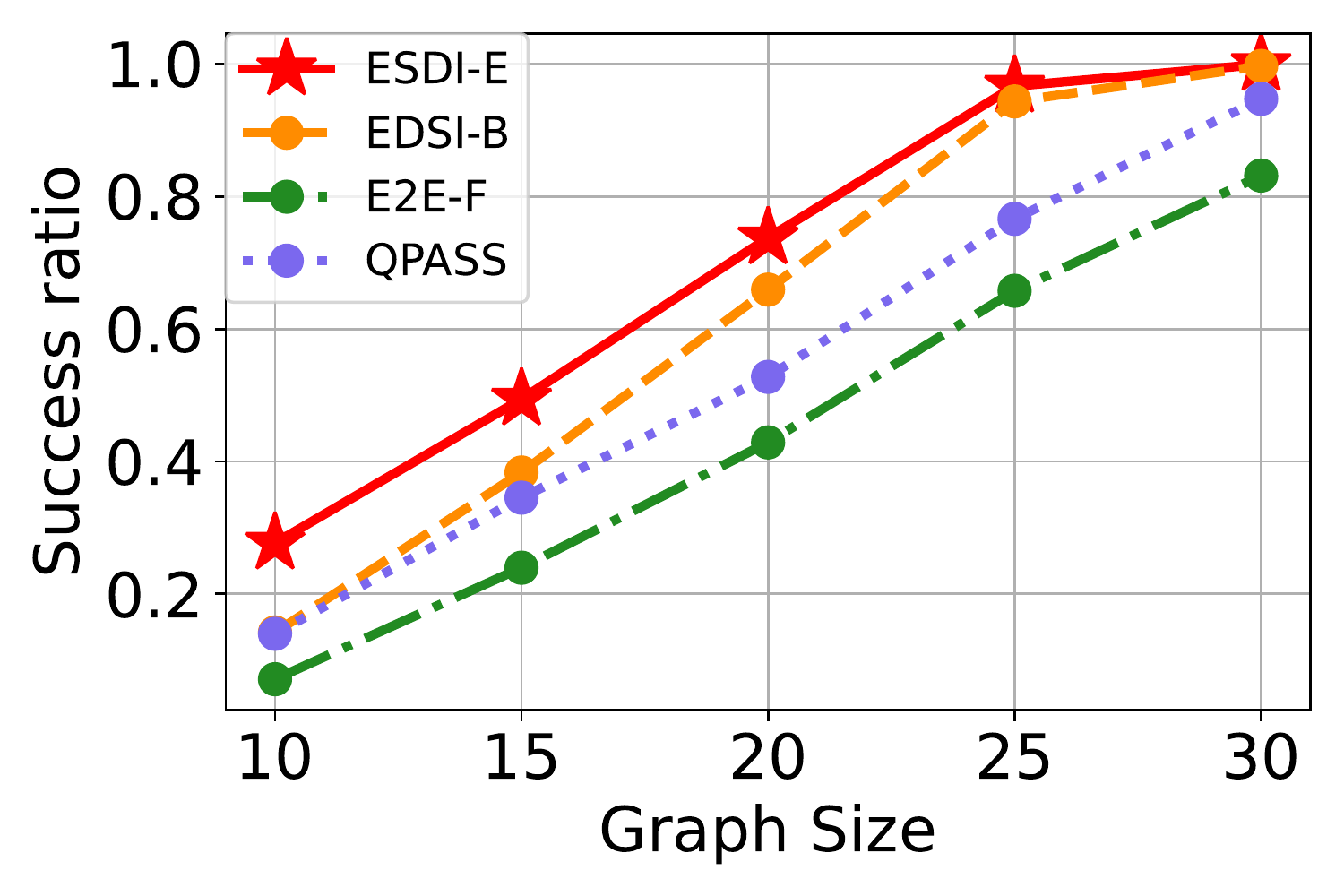}
\label{fig:algs_ddl:success_ratio_graph_ddl}}
\caption{Success ratio between ESDI and state-of-the-art algorithms}
\label{fig:algs_ddl}
\end{figure*}

\begin{figure*}[h]
\vspace{-1em}
\subfloat[Impact of the arrival rate]{\includegraphics[width=0.28\textwidth]{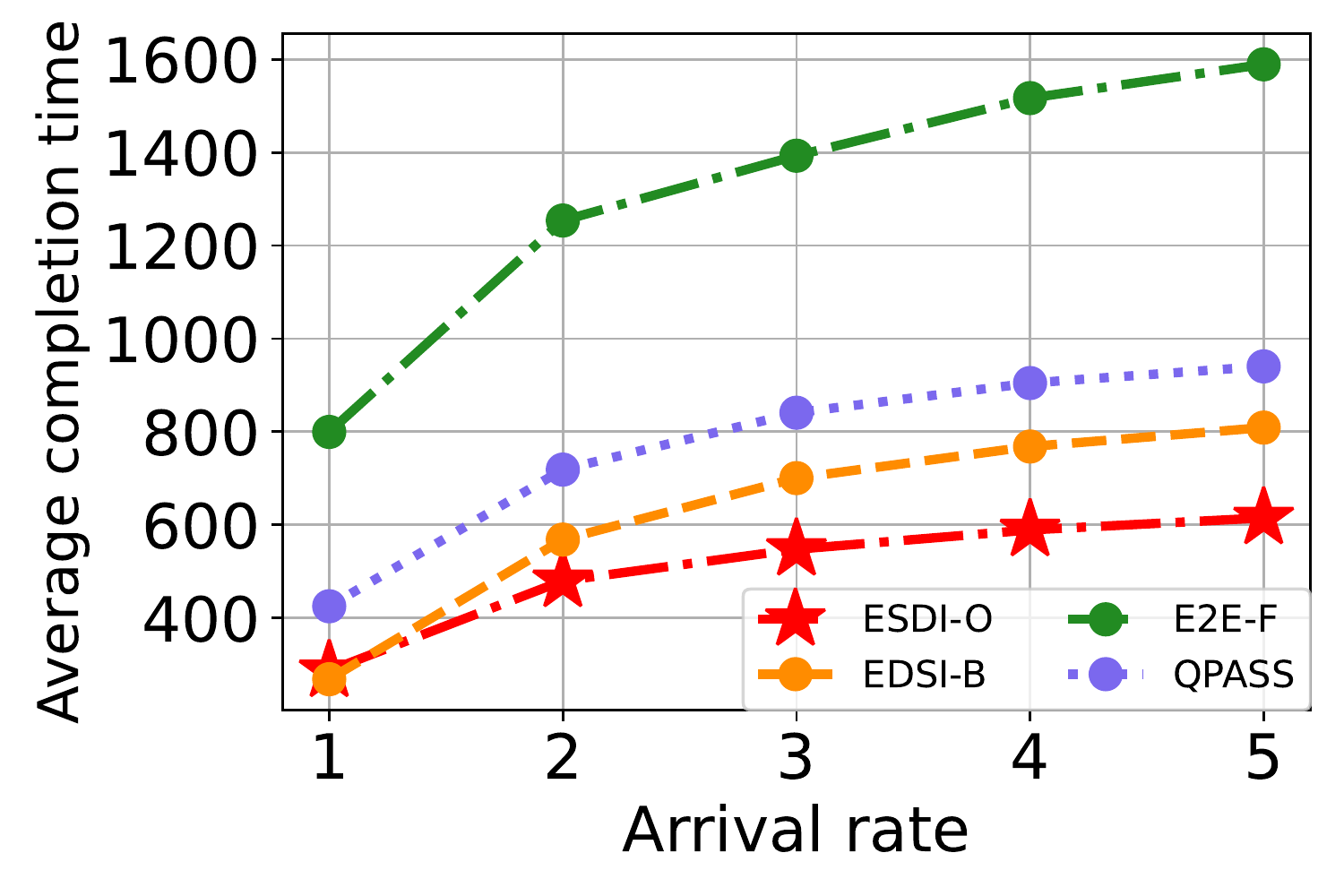}
\label{fig:algs_noddl:avg_comple_time_pos_noddl}}
\hfil
\subfloat[Impact of the demand]{\includegraphics[width=0.28\textwidth]{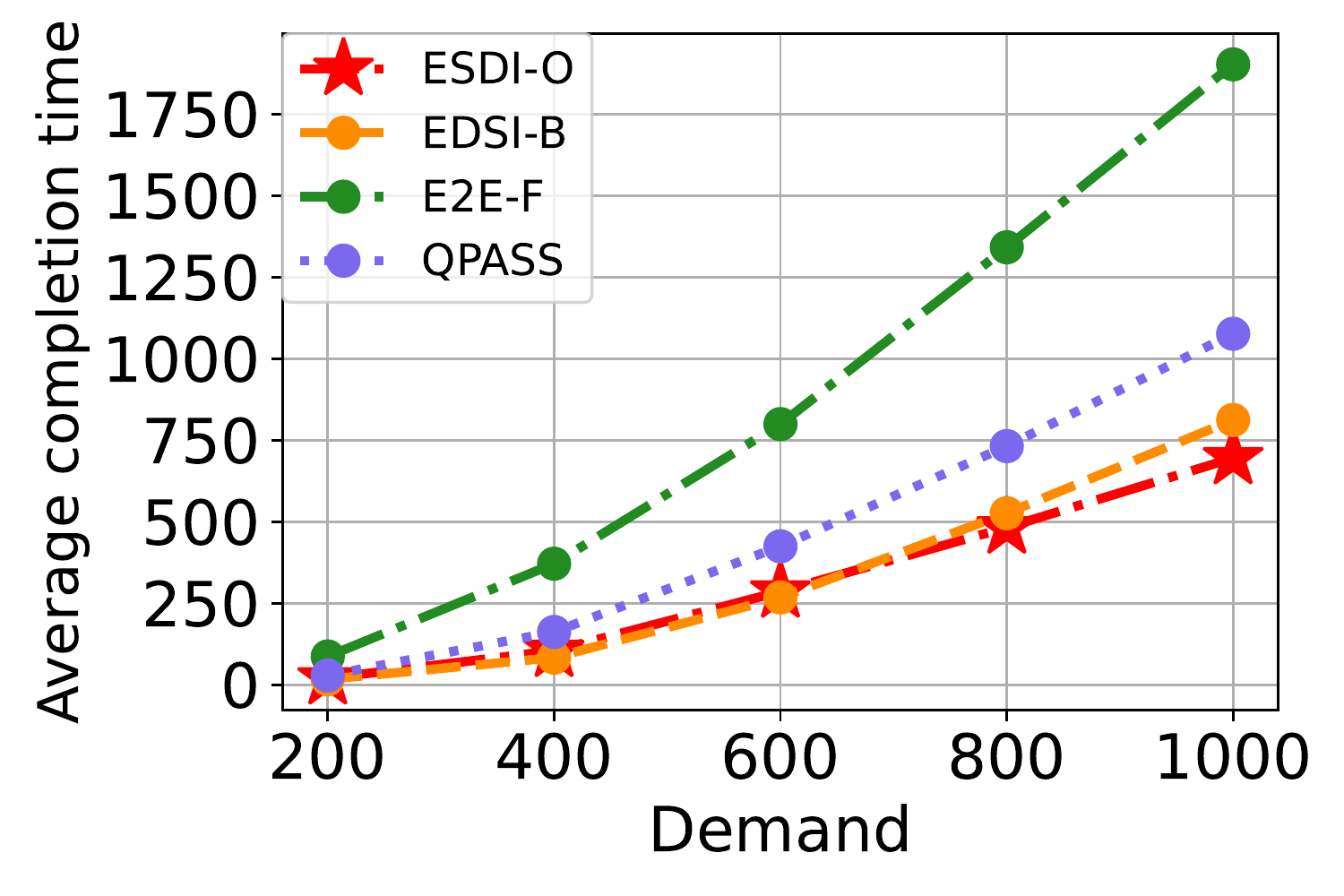}
\label{fig:algs_noddl:avg_comple_time_demand_noddl}}
\hfil
\subfloat[Impact of the graph size]{\includegraphics[width=0.28\textwidth]{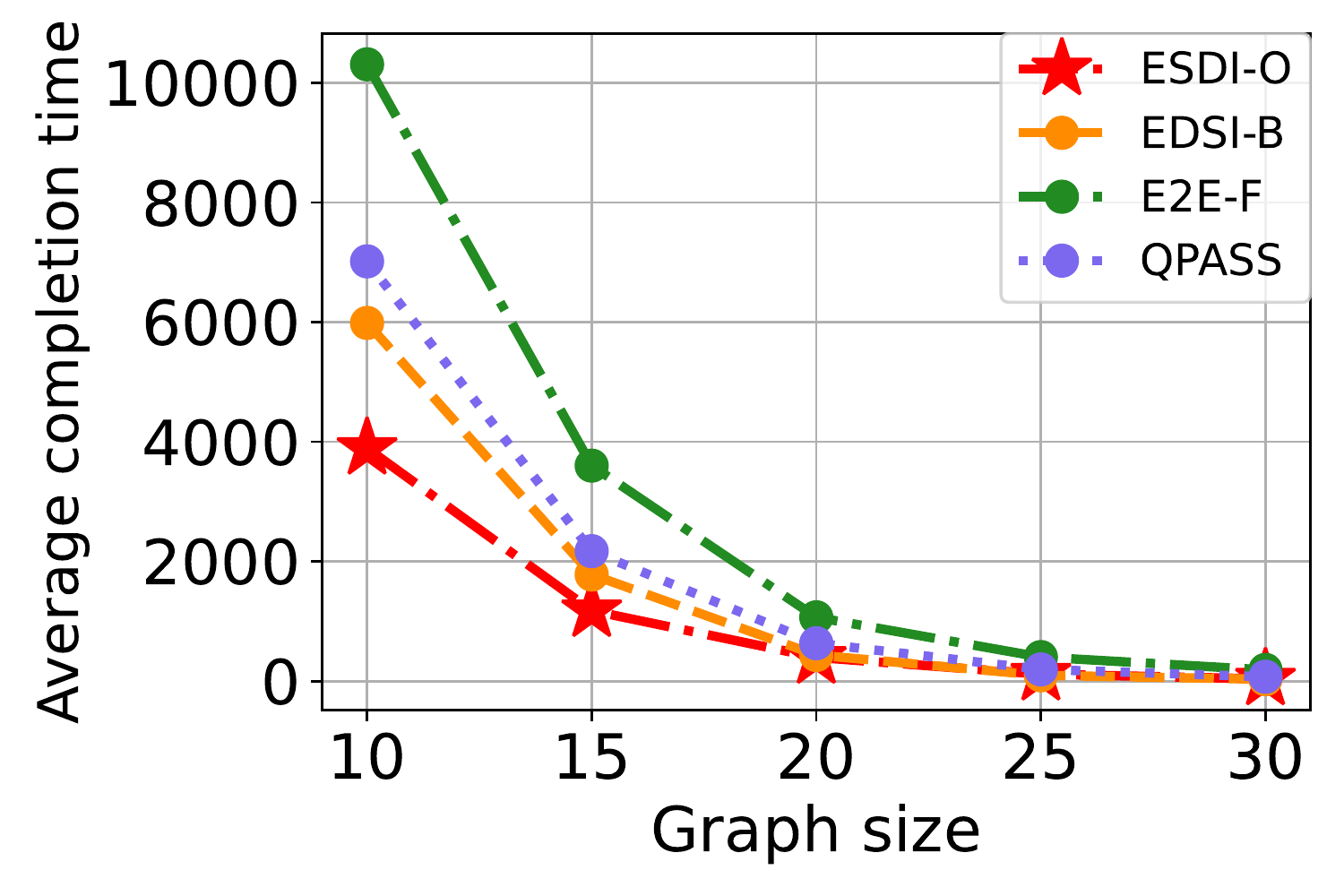}
\label{fig:algs_noddl:avg_comple_time_graph_noddl}}
\caption{Average completion time between ESDI and state-of-the-art algorithms}
\label{fig:algs_noddl}
\end{figure*}

\begin{figure}[h]
\centering
\subfloat{\includegraphics[width=0.48\textwidth]{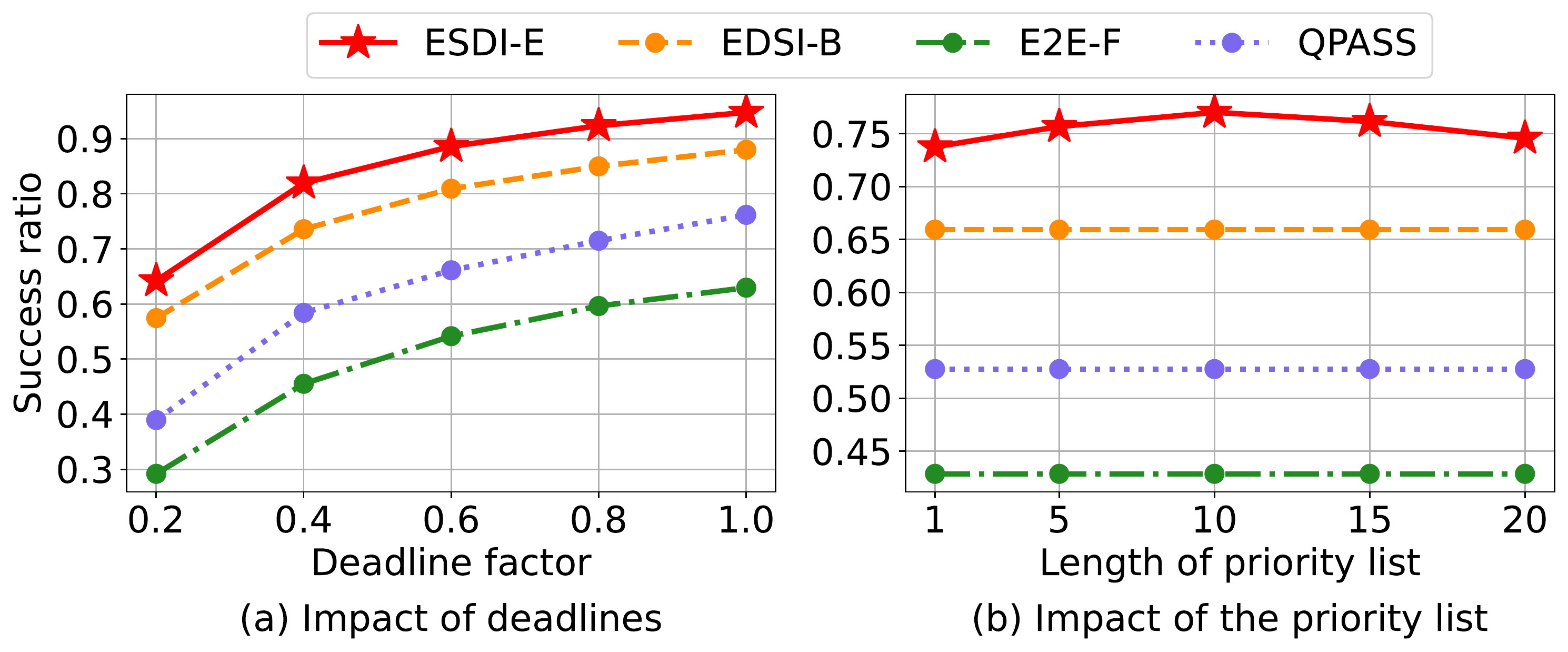}
\label{fig:alg_ddl}}
\caption{Impact of deadlines and the length of priority list}
\label{fig:algs_ddl:merge}
\end{figure}

\section{Performance Evaluation}
\label{sec:results}
\subsection{Evaluation Methodology}
\noindent
To evaluate our solution, we developed a discrete-time quantum network simulator.
We used random Waxman graphs~\cite{waxman} with model parameters $\alpha = \beta = 0.8$.
The success probability of entanglement generation and swapping were $0.9$.
Each link had a capacity uniformly sampled from $[3, 10]$.
We generated graphs with $20$ nodes and picked $1000$ SD pairs randomly.

Our simulator was based on a time-slotted model to be compatible with existing algorithms.
We implemented our ESDI protocol and two state-of-the-arts.
Linear programs were solved using the Gurobi solver~\cite{gurobi}.

The following algorithms were compared:
\begin{itemize}
    \item \textbf{ESDI-B:} basic ESDI without scheduling as in~\eqref{fml:0};
    \item \textbf{ESDI-O:}
    ESDI without deadlines in Algorithm~\ref{a:3};
    \item \textbf{ESDI-E:}
    ESDI with deadlines in Algorithm~\ref{a:1};
   \item \textbf{E2E-F:} 
   fidelity-aware protocol in~\cite{zhao2022e2e} maximizing end-to-end EDR. We set fidelity as $1$ since it is not considered. 
   \item \textbf{QPASS:} QPASS protocol in~\cite{shi2020concurrent} trying to maximize end-to-end EDR for multiple SD pairs. 
\end{itemize}
Commodities arrived following a Poisson distribution with arrival rate $\lambda = 1$ by default.
Demands of each commodity followed an exponential distribution with mean of $600$ ebits, and a minimum demand of $100$ ebits per commodity.
For a commodity with a deadline, we considered the difference between its deadline and arrival time to be a random function with expectation proportional to its demand.
Define $\mu = 0.4$. A unit deadline $\bar \delta^i_j$ was drawn from a uniform distribution in the range $[\mu - 0.1, \mu + 0.1]$ for each commodity, and the deadline of the commodity with demand  $d^i_j$ was set as $\delta^i_j = a^i_j + \bar \delta^i_j \cdot d^i_j$.
We set the default scheduling length $\kappa = 1$.
Both QPASS and E2E-F use Yen's algorithm where we set the number of paths to $15$.
Since they are \emph{entanglement routing} algorithms for bufferless quantum networks, we simulated their bufferless behavior by dropping all ebits after one time slot. 
Each simulation was run with $5$ seeds to reduce random noise.

The following metrics were used for evaluation.
\textbf{\emph{Success ratio}} measures the ratio of the number of commodities finished before their deadlines.
\textbf{\emph{Average completion time}} measures the average time between each commodity's arrival and completion when there is no deadline.

\subsection{Evaluation Results}

\subsubsection{Success ratio for commodities with deadline}
Fig.~\ref{fig:algs_ddl} shows the success ratio of ESDI-E and other algorithms with varying parameters, while all commodities have deadlines.
The success ratio decreased with increasing arrival rate in Fig.~\ref{fig:algs_ddl}\subref{fig:algs_ddl:success_ratio_pos_ddl} and increasing per-commodity demand in Fig.~\ref{fig:algs_ddl}\subref{fig:algs_ddl:success_ratio_demand_ddl} due to more severe resource contention, and increased with increasing graph size in Fig.~\ref{fig:algs_ddl}\subref{fig:algs_ddl:success_ratio_graph_ddl} due to more abundant resources in the network.
From all figures, ESDI-E and ESDI-B achieved the highest success ratio compared to other algorithms, and ESDI-E further improved success ratio by up to $55\%$ over ESDI-B.
This leads to two critical observations: 1) \emph{\textbf{MRED}} with optimal EDR in a buffered quantum network can significantly \emph{\textbf{improve network-wide throughput}} over existing heuristic-based bufferless algorithms, and 2) \emph{\textbf{scheduling} via prioritization} (ESDI-E) can additionally \emph{\textbf{finish more commodities before deadlines}}.

\subsubsection{Average completion time for commodities without deadline}
Fig.~\ref{fig:algs_noddl} shows the average completion time for ESDI-O and other algorithms with varying parameters. %
With higher arrival rate in Figs.~\ref{fig:algs_noddl}\subref{fig:algs_noddl:avg_comple_time_pos_noddl} or higher demand in Fig.~\ref{fig:algs_noddl}\subref{fig:algs_noddl:avg_comple_time_demand_noddl}, the average completion time of all algorithms increased, again due to more severe contention of resources.
Larger graph size decreased average completion time in Fig.~\ref{fig:algs_noddl}\subref{fig:algs_noddl:avg_comple_time_graph_noddl} by alleviating the contention.
The key observation across figures is again that ESDI-O and ESDI-B outperformed the other algorithms by 1) \emph{\textbf{maximizing network-wide throughput}} with MRED in ESDI-B, and 2) \emph{\textbf{scheduling} via prioritization} in ESDI-O.
\subsubsection{Impact of deadlines} 
Fig.~\ref{fig:algs_ddl:merge}(a) shows the success ratio for ESDI-E and other algorithms with varying deadline factors.
For each deadline factor value (such as $0.8$), we multiplied the deadline of every commodity with the factor.
From the figure, we observe increasing success ratios with increasing deadline factors, due to more commodities being able to finish before their deadlines.
The advantage of ESDI-E (and ESDI-B compared to other heuristics) was consistent across different deadlines, showing the constant advantage of our algorithms.

\subsubsection{Impact of scheduling length $\kappa$}
Fig.~\ref{fig:algs_ddl:merge}(b) shows the success ratio for ESDI-E and other algorithms with varying lengths of the priority list for scheduling. Because there was no priority list in ESDI-B, E2E-F, and QPASS, their success ratios kept the same.
For ESDI-E, we observed a trade-off between \emph{scheduling} and \emph{work conservation}, the two primary goals we defined in \S\ref{sec:general_framework}.
Specifically, increasing the scheduling length $\kappa$ from the minimum value $1$ to an intermediate value $10$ slightly increased the success ratio of ESDI-E, while further increasing $\kappa$ led to decrease in the success ratio.
The initial increase was due to our algorithm being able to \emph{prioritize} commodities without significant impact on the throughput of other commodities in the network.
As prioritization increased further, throughput started to degrade since some prioritized commodities (those with earlier deadlines) might span across longer paths, thus having lower end-to-end EDR than other commodities with later deadlines but shorter paths.
In practice, the exact optimal length $\kappa$ depends on the network topology and the set of commodities, and finding this optimal scheduling length is an important future work.
Fortunately, as we can observe, even setting $\kappa = 1$ (scheduling only one commodity at a time) still led to a substantial advantage over any non-scheduling baseline, thus motivating us to set $\kappa=1$ as the default value for evaluation.

\section{Conclusions}
\label{sec:conclusions}
\noindent 
In this paper, we designed a general optimization framework \textbf{ESDI} for entanglement scheduling and distribution in a general quantum network.
Motivated by scenarios where different quantum communication applications have different demands and time requirements (such as deadlines), we first developed a multi-commodity entanglement distribution formulation, and then designed two scheduling and distribution algorithms based on the idea of \emph{scheduling via prioritization}.
We further designed a practical probabilistic protocol to implement the optimized ESDI decisions in a buffered quantum network.
We developed a discrete-time quantum network simulator for evaluation.
Extensive simulation results showed our solutions could significantly reduce the average completion time of quantum communication demands and increase the success ratio of commodities when compared to existing entanglement routing and distribution algorithms.
%
%
%
%
%
%


%
%
%

%
%
%
%
%
%
\end{document}